\documentclass{article}
 \pdfoutput=1
\usepackage{url}
\usepackage{newpxtext,newpxmath}

\usepackage{amsmath, amssymb}
\usepackage{fullpage}
\usepackage{natbib}
\usepackage{graphicx}
\usepackage[frozencache,cachedir=.]{minted}
\usepackage{xcolor}
\usepackage{authblk}
\usepackage{xr-hyper}
\usepackage{hyperref}
\hypersetup{colorlinks=true,
        urlcolor=violet,
	linkcolor=teal,
	citecolor=black}

\newcommand\mbb[1]{\mathbb{#1}}

\def\absarg#1{\left|#1\right|}

\def\reals{\mathbb{R}} % Real number symbol
\def\*#1{\mathbf{#1}}

\def\PD{\mathrm{PD}}
\def\FDP{\mathrm{FDP}}
\renewcommand{\exp}[1]{\operatorname{exp}\left(#1\right)} % Exponential
\def\indic#1{\mbb{I}\left({#1}\right)} % Indicator function

\makeatletter
\newcommand*{\addFileDependency}[1]{% argument=file name and extension
\typeout{(#1)}
\@addtofilelist{#1}

\IfFileExists{#1}{}{\typeout{No file #1.}}
}\makeatother

\begin{document}

% Title of paper
\title{mbtransfer: Microbiome Intervention Analysis using Transfer Functions and Mirror Statistics}

% List of authors, with corresponding author marked by asterisk
%\author{KRIS SANKARAN$^{\ast,1,2}$, PRATHEEPA JEGANATHAN$^{3}$\\[4pt]
% Author addresses
%\textit{$^{\ast,1}$ Department of Statistics, University of Wisconsin - Madison,
%Department of Statistics,
%University of Wisconsin
%1300 University Ave
%Madison, WI 53706
%USA
%}
%\\[2pt]
%\textit{$^{2}$ Wisconsin Institute for Discovery
%}
%\\[2pt]
%\textit{$^{2}$ Department of Mathematics \& Statistics, McMaster University
%}
%\\[2pt]
% E-mail address for correspondence
%{ksankaran@wisc.edu}
%}

 \author[1,2]{Kris Sankaran}
 \author[3]{Pratheepa Jeganathan}
 \affil[1]{Department of Statistics, University of Wisconsin - Madison}
 \affil[2]{Wisconsin Institute for Discovery}
 \affil[3]{Department of Mathematics \& Statistics, McMaster University}

%\begin{document}

% Running headers of paper:
\markboth%
% First field is the short list of authors
{K. Sankaran and P. Jeganathan}
% Second field is the short title of the paper
{Microbiome Intervention Analysis with Transfer Functions and Mirror Statistics}
\maketitle

\begin{abstract}
{Microbiome interventions provide valuable data about microbial ecosystem structure and dynamics. Despite their ubiquity in microbiome research, few rigorous data analysis approaches are available. In this study, we extend transfer function-based intervention analysis to the microbiome setting, drawing from advances in statistical learning and selective inference. Our proposal supports the simulation of hypothetical intervention trajectories and False Discovery Rate-guaranteed selection of significantly perturbed taxa. We explore the properties of our approach through simulation and re-analyze three contrasting microbiome studies. An R package, mbtransfer, is available at \href{https://go.wisc.edu/crj6k6}{https://go.wisc.edu/crj6k6/}. Notebooks to reproduce the simulation and case studies can be found at \href{https://go.wisc.edu/dxuibh}{https://go.wisc.edu/dxuibh} and \href{https://go.wisc.edu/emxv33}{https://go.wisc.edu/emxv33}.}
%key words 
%{Intervention analysis, microbiome dynamics, longitudinal data}
\end{abstract}

\section{Introduction}
\label{sec1}

\begin{figure}[!p]
\centering\includegraphics[width=\textwidth]{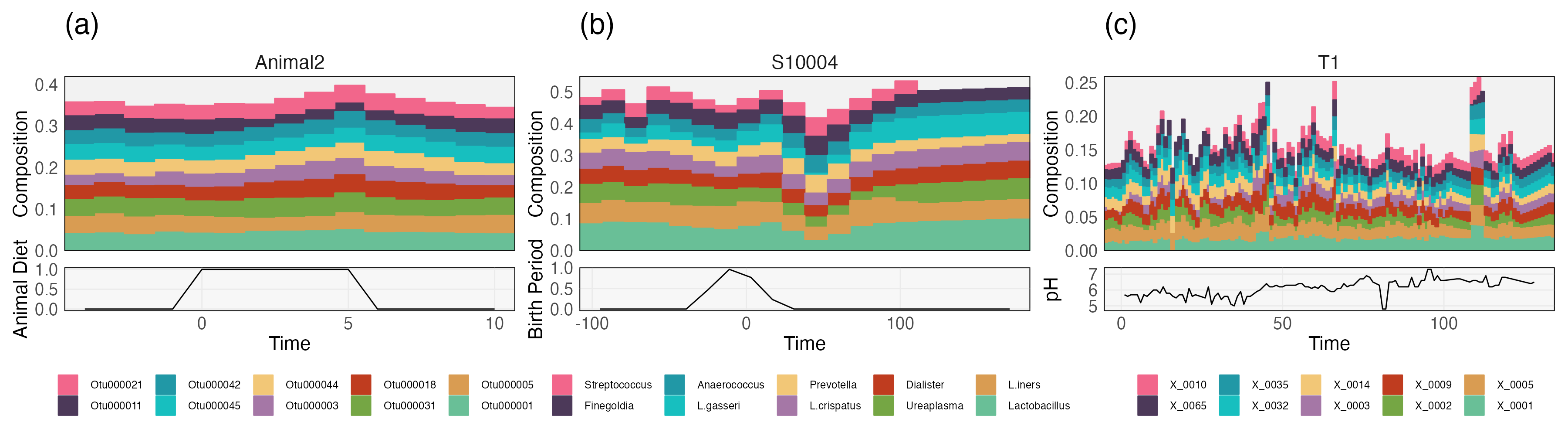}
\caption{Examples of microbial community shifts in response to environmental change. Part (a) describes the gut microbiome of a subject undergoing a diet intervention \citep{David2013DietRA}, (b) shows remodeling of a mother's  vaginal microbiome following birth \citep{Costello2022LongitudinalDO}, and (c) profiles an aquaculture tank microbiome together with environmental pH \citep{Yajima2022CoreSA}. Section \ref{sec:data_analysis} explores data from these studies in depth.}
\label{fig:opening}
\end{figure}

Figure \ref{fig:opening} gives three examples of microbial community dynamics under environmental perturbations. Part (a) shows the shift in the gut microbiome of a subject from \citet{David2013DietRA} during a five-day shift to an animal-based diet. Part (b) shows the postpartum change in the vaginal microbiome from one participant tracked by \citet{Costello2022LongitudinalDO}. Part (c) gives the dynamics of an aquaculture microbiome in a tank following shifts in environmental pH, as examined by \citet{Yajima2022CoreSA}. 
The shifts in these few cases represent general phenomena -- the interventions they describe have reproducible effects on the microbiome, consistently altering the abundance of specific taxa on a predictable time scale. Similar microbial community studies are widespread in microbiome research efforts, especially those with the long-term goal of engineering microbial systems that promote health in a dynamic environment.

Statistical inference of intervention effects on the microbiome must account for temporal dependence -- otherwise, there is a risk of overinflating the effective sample size. We will see in Section \ref{sec:simulations} that, though the practice of two-sample testing of intervention effects is common, it leads to inflated false discovery rates. Several microbial community dynamics models have been proposed in response to this issue. Among the most widely used is the generalized Lotka-Volterra (gLV) model, which discretizes an ordinary differential equation model for competitive predator-prey dynamics, optionally including covariates to model environmental influences \citep{Gerber2014TheDM, Gibbons2017TwoDR}. Specifically, let $\*y\left(t\right)$ be the microbial community profile at time $t$, and let $\*w\left(t\right)$ be the state of the associated intervention.
Then, the gLV supposes $\frac{\partial\*y\left(t\right)}{\partial t} = A\*y\left(t\right) + D\*w\left(t\right) + \epsilon\left(t\right)$, and it is typically estimated by first log-transforming the observed taxonomic abundances $\log\left(1 + \*y_{t}\right)$ and fitting an elastic net regression of $\log\left(1 + \*y_{t + 1}\right) - \log\left(1 + \*y_{t}\right)$ onto $\*w_{t}$. The main limitations of this model are (1) that it assumes linearity in the relationship of $\log \left(1 + \*y_{t + 1}\right) - \log\left(1 + \*y_{t}\right)$ onto $\*w_{t}$ and (2) it can only refer to the immediate past of and $\*w_{t}$. Moreover, it does not quantify the uncertainty or stability of any estimated effects.

To address this, \citet{Bucci2016MDSINEMD, Gibson2021IntrinsicIO} and \citet{Silverman2018DynamicLM, silverman2022} developed explicit probabilistic models of community dynamics. \citet{silverman2022}'s models, MALLARD and fido, are based on a multinomial logistic-normal autoregressive and multinomial logistic-normal Gaussian Process model, respectively. The environmental shifts can be included as covariates to support intervention analysis. MDSINE and MDSINE2 \citep{Bucci2016MDSINEMD, Gibson2021IntrinsicIO} are negative binomial dynamical systems models that extend the gLV and focus on the discovery of interspecies interactions and perturbation effects. Autoregressive dynamics are clustered using a Dirichlet Process, and a Gaussian Process prior is used to regularize species abundance trajectories. These models are closely related to our work in their application of a dynamical systems model and inference of environmental intervention effects. We provide a quantitative comparison in Section \ref{sec:simulations}, and Supplementary Section \ref{subsec:reference_methods} summarizes existing methods.

We make two contributions to the tapestry of currently available models. First, we demonstrate that nonparametric transfer function models lead to more accurate forecasts than current models, especially when environmental shifts are large. We leverage an existing gradient boosting package \citep{Chen2018GMPRAR} and achieve competitive performance, most likely due to their relatively weak modeling assumptions and our data-rich setting. Second, we provide an algorithm to support the precise inference of intervention effects on individual taxa at specific temporal lags. Decoupling community dynamical modeling from inference makes our interpretations of environmental effects more robust to model misspecification. Section \ref{sec:method} explains how to guarantee False Discovery Rate (FDR) control of the selected taxa using only a symmetry assumption.

The key ingredients of our approach are transfer function models \citep{Box1975InterventionAW}, which summarize community dynamics, and mirror statistics \citep{Dai2020FalseDR}, which enable precise inference. Transfer functions relate an ``input'' series to an ``output'' one. These models were originally developed to support intervention analysis in time series data, for example, the influence of a new automobile emissions regulation on local ozone levels. Section \ref{sec:method} adapts this framework to the high-dimensional microbiome setting. Mirror statistics are an approach to selective inference that leverages data splitting to rank differential effects while controlling the FDR, and we develop an instance of this algorithm using partial dependence profiles of the fitted boosting models. This approach is analogous to recent microbiome approaches based on knockoffs \citep{Xie2021AggregatingKF, Zhu2021DeepLINKDL}, but it does not depend on the simulation of appropriate knockoff features.

Taken together, transfer function models and mirror-based inference provide answers to the following central questions in microbiome intervention analysis:
\begin{enumerate}
\item Which taxa are the most affected by the intervention? Our mirror statistics identify taxa with differential trajectories across counterfactual environmental conditions.
\item When are these taxa affected? We can distinguish between transient and sustained shifts in the community by simulating alternative counterfactuals from our fitted transfer function models.
\item Which factors mediate the shift? Flexible transfer function models can detect interactions between interventions and environmental features.
\end{enumerate}

The mbtransfer R package computes artifacts directly related to these questions. Specifically, it supports training transfer function models, testing for significant taxon-level effects, and simulation under counterfactual alternatives. The package's implementation and documentation, including the code to reproduce the data analysis of Section \ref{sec:data_analysis}, can be found at \href{https://go.wisc.edu/crj6k6}{https://go.wisc.edu/crj6k6}.

\section{Method}
\label{sec:method}

Section \ref{subsec:transfer_function} discusses a flexible generalization of transfer function models \citep{Box1975InterventionAW}. Here, the input series is a measure of the intervention strength. The resulting model can summarize and simulate intervention effects on microbiome communities, accounting for baseline composition and mediating host features. Section \ref{sec:mirror_statistics} develops mirror statistics \citep{Dai2020FalseDR} to formally infer which taxa are the most strongly influenced by the interventions and whether the effects are immediate or delayed. The overall workflow supports statistically-guaranteed discovery of intervention effects in microbial time series.

\subsection{Transfer function models}
\label{subsec:transfer_function}

Transfer function models were introduced as a linear autoregressive model applied to two concurrent time series, a series $w_{t} \in \reals$ that serves as the intervention and a series $y_{t} \in \reals$ that changes in response. We consider the generalization,
\begin{align}
	y_{tj}^{(i)} &= f_{j}\left(\*y_{(t - P - 1):(t - 1)}^{(i)}, \*w_{(t - Q + 1):t}^{(i)}, \*z^{(i)}\right) + \epsilon^{(i)}_{jt},
\label{eq:transfer_model}
\end{align}
where we have used the following notation:
\begin{itemize}
	\item $\*y_{t}^{(i)} \in \reals^{J}$: The (potentially transformed) abundances of all taxa $j \in \{1, \dots, J\}$ at time $t$ in subject $i$. This vector has $j$th coordinate $y_{tj}^{(i)}$.
	\item $\*w_{t}^{(i)} \in \reals^{D}$: The strength of $D$ different interventions at time $t$ in subject $i$.
	\item $\*z^{(i)} \in \reals^{S}$: The characteristics of subject $i$ that do not vary over time.
	\item $\*\epsilon_{jt}^{(i)}$: Random error in the abundance of taxon $j$ for timepoint $t$ in subject $i$. In Section \ref{sec:mirror_statistics}, this noise is assumed symmetric.
\end{itemize}
We learn each $f_{j}$ separately using gradient boosting models \citep{Friedman2001GreedyFA,Chen2016XGBoostAS}. For training, we extract nonoverlapping temporal segments, and the last $P$ lags of $\*y_{t}^{(i)}$ and $Q$ lags of $\*w_{t + 1}^{(i)}$ are vectorized and combined with $\*z^{(i)}$ to form a $\reals^{PJ + QD + S}$-dimensional feature vector. Once trained, this model can simulate expected counterfactual trajectories under hypothetical interventions $\tilde{\*w}_{(t + 1):(t + h)}$ given initial compositions $\*y_{(t - P + 1):t}$ and subject characteristics $\tilde{\*z}$. The one-step forecast is as follows:
\begin{align*}
	\hat{\*f}\left(\*y_{(t - P + 1):t}, \tilde{\*w}_{(t - Q + 2):(t + 1)}, \tilde{\*z}\right) &:= \begin{bmatrix}
		\hat{f}_{1}\left(\*y_{(t - P + 1):t}, \tilde{\*w}_{(t - Q + 2):(t + 1)}, \tilde{\*z}\right) \\
		\vdots \\
		\hat{f}_{J}\left(\*y_{(t - P + 1):t}, \tilde{\*w}_{(t - Q + 2):(t + 1)}, \tilde{\*z}\right)
	\end{bmatrix}
\end{align*}
and longer time horizons can be forecast by substituting intermediate predictions:
\begin{align*}
\hat{\*f}^{+h}\left(\hat{\*y}_{(t - P + h):(t + h - 1)}, \tilde{\*w}_{(t - Q + h + 1):(t + h)}, \tilde{\*z}\right) &:= \begin{bmatrix}
	\hat{f}_{1}\left(\hat{\*y}_{(t - P + h):(t + h - 1)}, \tilde{\*w}_{(t - Q + h  + 1):(t + h)}, \tilde{\*z}\right) \\
	\vdots \\
	\hat{f}_{J}\left(\hat{\*y}_{(t - P + h):(t + h - 1)}, \tilde{\*w}_{(t - Q + h + 1):(t + h)}, \tilde{\*z}\right)
\end{bmatrix} 
\end{align*}
where we used the convention that $\hat{\*y}_{t'} = \*y_{t'}$ if $t' \leq t$ is observed and $\hat{\*y}_{t'} = \hat{\*f}^{+h'}\left(\hat{\*y}_{(t - P + h'):(t + h' - 1)}, \tilde{\*w}_{(t - Q + h' + 1):(t + h')}, \tilde{\*z}\right)$ for intermediate predictions $t' = t + h'$ with $ h' < h$.

The two advantages of this formulation are: (1) it can estimate nonlinear relationships between past microbial community profiles, interventions, and host features with taxon $j$'s current abundance, and (2) it can detect interaction effects between covariates that improve predictive power and which may have valuable scientific interpretations. Note that each taxon $j$ is trained separately. On the one hand, this means that information is not shared between related taxa. On the other, this allows us to use existing, reliable boosting implementations, and if many taxa are of interest, the implementation is easily parallelizable. 

%\subsection{Identify Transfer Function (XGboost) and Diagnostics}

%Stability analysis of fitted transfer function models or model identification, specifically the response function. We group the taxa based on the estimated transfer function weights and mirror statistic-based testing. 

\subsection{Mirror statistics}
\label{sec:mirror_statistics}

The transfer function model in equation (\ref{eq:transfer_model}) summarizes the effects of interventions $\*w_{t}$ on taxonomic abundances $\*y_{t}$. However, this model may not provide statistical guarantees and can lead to ambiguous results. To address this, we propose a mirror statistics implementation. First, we randomly split subjects into subsets, $\mathcal{D}^{(1)}$ and $\mathcal{D}^{(2)}$. Then, for each split $s$, we train models $\hat{\*f}^{(s)}$. Next, we estimate the counterfactual difference between interventions $\tilde{\*w}_{(t - Q + 2):(t + 1)} = \*1_{Q}$ and $\tilde{\*w}_{(t - Q + 2):(t + 1)} = \*0_{Q}$ for each taxon $j$ using:
\begin{align}
	\PD_{j}^{(s)} &= \frac{1}{\absarg{\mathcal{D}^{(s)}}}\sum_{\text{segments} \in \mathcal{D}^{(s)}}\left[\hat{f}_{j}^{(s)}\left(\*y^{(i)}_{(t - P + 1):t}, \*1_{Q}, \*z^{(i)}\right) - \hat{f}_{j}^{(s)}\left(\*y^{(i)}_{(t - P + 1):t}, \*0_{Q}, \*z^{(i)}\right)\right].
\end{align} 
This equation is a partial dependence profile applied to the fitted model of taxon $j$ \citep{Friedman2001GreedyFA, Biecek2021ExplanatoryMA}. Note that this definition toggles all $D$ interventions. For an isolated intervention, we can use $\left(0, \dots, 0, 1, 0, \dots, 0\right)$ instead of $\*1_{Q}$ to get the analogous statistic. We also define the corresponding mirror statistic as:
\begin{align}
	M_{j} = \text{sign}\left(\PD_{j}^{(1)}\PD_{j}^{(2)}\right)\left[\absarg{\PD_{j}^{(1)}} + \absarg{\PD_{j}^{(2)}}\right],
\end{align}
which measures the consistency between estimated effects across separate splits. We assume that for taxon $j$ with no true intervention effects, $\PD_{j}^{(s)}$ is symmetrically distributed around 0. This assumption is plausible because in the absence of an intervention effect on taxon $j$, any differences between $\hat{f}_{j}^{(s)}\left(\*y^{(i)}_{(t - P + 1):t}, \*1_{Q}, \*z^{(i)}\right)$ and $\hat{f}_{j}^{(s)}\left(\*y^{(i)}_{(t - P + 1):t}, \*0_{Q}, \*z^{(i)}\right)$ are due to noise.

Given mirror statistics $M_{j}$, we estimate the false discovery proportion using the same procedure of \cite{Dai2020FalseDR}, viewing $\PD^{(s)}_{j}$ as analogous to $\hat{\beta}_{j}^{(s)}$. Specifically, we compute:
\begin{align}
\label{eq:fdp_est}
	\widehat{\FDP}\left(t\right) &= \frac{\absarg{\{j : M_{j} < -t\}}}{\absarg{\{j : M_{j} > t\}}},
\end{align}
where the choice of $t$ defines a selection set $\hat{J}_{1}$ for the current pair of splits. Given FDR control level $q$, we choose the largest $t$ such that $\widehat{\FDP}\left(t\right) \leq q$. We aggregate across multiple splits to improve power, following \citet{Dai2020FalseDR}'s Algorithm 2. Our examples always aggregate across 25 splits. For delayed effects, we define analogous $\PD_{j}^{(s),+h}$ and $M_{j}^{+h}$ using $f_{j}^{+h}$ instead of $f_{j}$, and the estimate in equation \ref{eq:fdp_est} is modified to use mirrors across all lags $h$.

\section{Simulations}
\label{sec:simulations}
We perform simulation experiments to examine the effectiveness of mbtransfer relative to competitive baselines and to identify data characteristics that lead to better or worse performance. We evaluate both forecasting ability and inferential quality.

\subsection{Data generating mechanism}
\label{sec:data_generating_mechanism}

We simulate data from a negative binomial vector autoregressive model:
\begin{align*}
\*y_{t}^{(i)} \vert \theta_{t}^{(i)}, \*\varphi_{i} &\sim \text{NB}\left(\exp{\theta_{t}^{(i)}}, \*\varphi_{i}\right) \\
\theta_{t}^{(i)} \vert \epsilon_{t}^{(i)} &= \sum_{p = 1}^{P} A_{p}\theta_{t - p}^{(i)} + \sum_{q = 1}^{Q} \left(B_{q} + C_{q} \odot z^{i}\right) \*w_{t - q}^{i} + \*\epsilon_{t}^{(i)}\\
\varphi_{ij} &\sim \Gamma\left(\alpha, \lambda\right) \\
\epsilon_{t}^{(i)} &\sim \mathcal{N}\left(0, \sigma^2_{\epsilon} I_{J}\right)
\end{align*}
Here, $\theta_{t}^{(i)} \in \reals^{J}$, and $\text{NB}$ refers to a negative binomial distribution applied coordinate-wise to each taxon $j$ using a mean-dispersion parameterization. $A_{p} \in \reals^{J \times J}$ parameterizes the lag-$p$ autoregressive dynamics between pairs of taxa and $B_{q} \in \reals^{J \times D}$ parameterizes the lag-$q$ effect of the $D$ interventions. We have chosen a negative binomial generative mechanism because this distribution has previously been found to fit 16S rRNA gene sequencing data well, especially after correcting for library size differences \citep{Calgaro2020AssessmentOS}. $C_{q}$ represents an interaction between host characteristics and the intervention, where some taxa may be more strongly affected by an intervention when their host has particular features.

The parameters $B_{q}, C_{q}$, and $A_{p}$ are simulated as follows. The first $J_{1}$ taxa have true intervention effects and the remaining $J_{0} = J - J_{1}$ rows of $B_{q}$ and $C_{q}$ are set to 0. Among the nonnull taxa $j \in J_{1}$, we draw $B_{q,jd} \sim  \text{Unif}\left(\left[-2b, -b\right]\cup \left[b, 2b\right]\right)$ where $b$ encodes the signal strength. Using two intervals ensures that nonnull effects are bounded away from 0. Entries $C_{q, jd}$ are drawn similarly, except entire rows $C_{q, j\cdot}$ are set to 0 with an additional probability $p_{c}$. Such rows represent taxa with real intervention effects but no interaction with host characteristics, represented by $z^{(i)} \sim \mathcal{N}\left(0, \sigma_{z}^{2}\right)$. Finally, we simulate $A \in \reals^{J \times J}$ as a sparsified version of a random, low-rank matrix. Specifically, we first set $\tilde{A}^{(0)} \sim QQ^{T}$ where $Q \in \reals^{J \times K}$ has entries drawn independently from $\mathcal{N}\left(0, \sigma^{2}_{A}\right)$. Entries of $\tilde{A}^{(0)}$ are randomly set to 0 with probability $p_{A}$, yielding $\tilde{A}^{(1)}$, and the result is normalized: $A = \frac{\tilde{A}^{(1)}}{\|\tilde{A}^{(1)}\|_{2}}$.

\begin{figure}[!p]
\centering\includegraphics[width=0.9\textwidth]{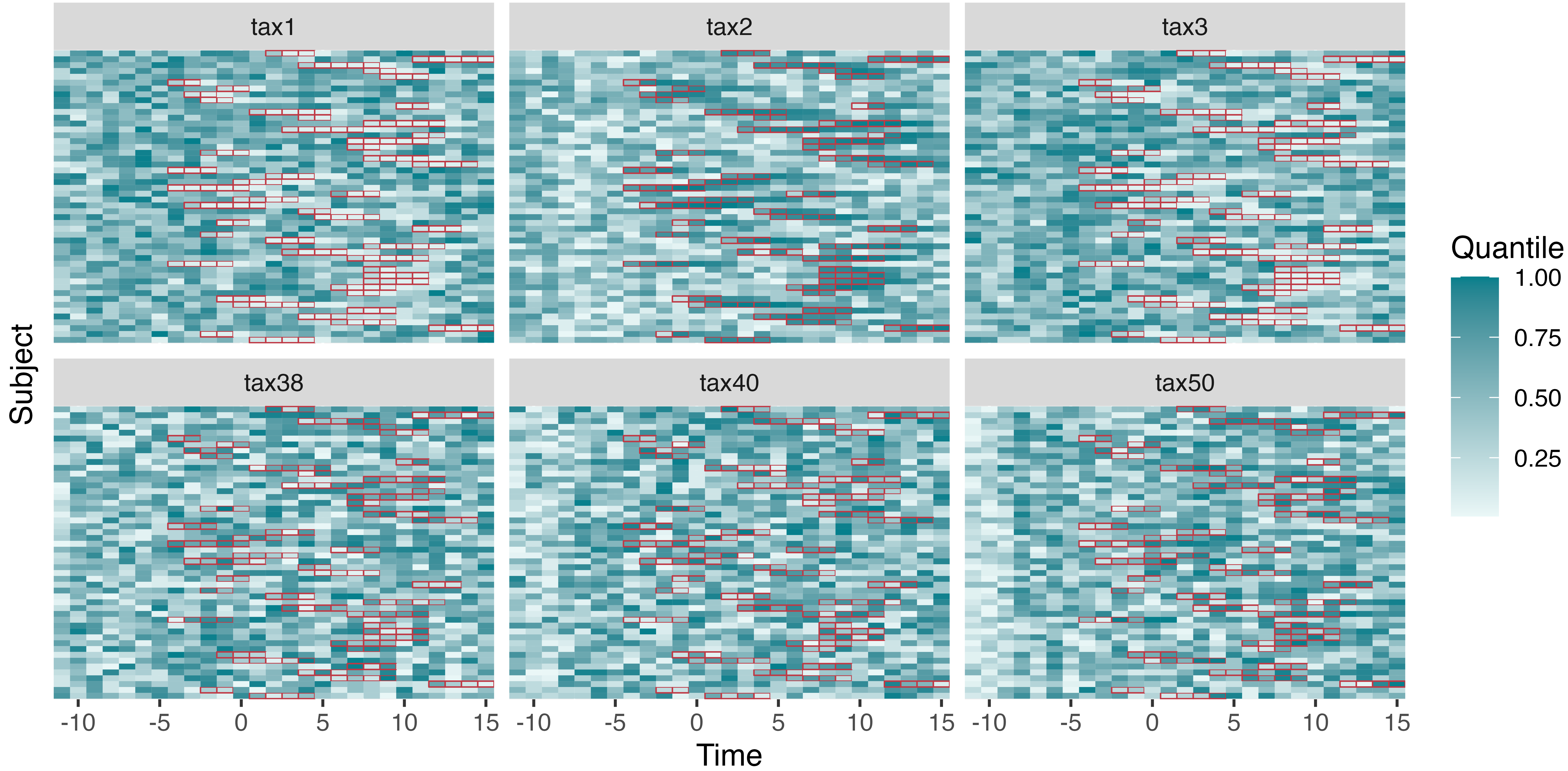}
	\caption{Time series simulated according to the mechanism described in Section \ref{sec:data_generating_mechanism}. Each panel shows the trajectories for one taxon, and each row is the time series for one subject. The colors of the tiles encode changes in abundance. Red borders indicate the samples where the intervention is present. The first row of taxa (tax1, tax2, tax3) are nonnull with negative, positive, and negative effects, respectively, and the bottom row of taxa are all null.}
	\label{fig:simulation-hm}
\end{figure}

We simulate random, one-dimensional interventions $w_{t}^{(i)} \in \{0, 1\}$ by first randomly sampling a starting point $t^{\text{start}} \sim \text{Unif}\left[\frac{T}{3}, \dots, \frac{2T}{3}\right]$. The intervention length is drawn from $\ell \in \text{Unif}\left[L, 2L\right]$. If $t^{\text{start}} + \ell > T$, we truncate the intervention series at $T$. A visualization of the trajectories for null and nonnull taxa is given in Figure \ref{fig:simulation-hm}. For inference, we compare the mirror algorithm to DESeq2 \citep{Love2014ModeratedEO} with the formula $\sim \texttt{intervention} + z^{(i)} + z^{(i)} \times \texttt{intervention}$. DESeq2 is a negative binomial-based generalized linear model originally developed for hypothesis testing in bulk RNA-seq data. Nonetheless, it is often recommended in 16S sequencing analysis and has exhibited strong performance in benchmarks against methods specifically built for 16S gene sequencing data \citep{Callahan2016BioconductorWF, Calgaro2020AssessmentOS}. Supplementary Section \ref{subsec:reproducibility} provides further details for examining and reproducing the simulation setup.

Given this simulation mechanism, we vary the following data parameters:
\begin{enumerate}
	\item The number of taxa $\in \left\{100, 200,400\right\}$.
	\item The fraction of null taxa $\pi_{0} \in \{0.1, 0.2, 0.4\}$. Given $\pi_{0}$, we set $J_{0} = \lfloor \pi_{0} J \rfloor$.
	\item The signal strength $b \in \{0.25, 0.5, 1\}$.
\end{enumerate}

Across all runs, we simulate 50 subjects with 30 timepoints each. We fix $p_{c} = 0.2$ and $p_{A} = 0.4$. Considering all combinations of these parameters yields 27 simulated datasets. These can be downloaded from \href{https://go.wisc.edu/8ey754}{https://go.wisc.edu/8ey754}, and simulation outputs discussed below are available at\footnote{Reproducibility details are discussed in Supplementary Section \ref{subsec:reproducibility}. Results were obtained using \citet{{chtc}}.} \href{https://go.wisc.edu/3gc982}{https://go.wisc.edu/3gc982}.

\subsection{Model settings and metrics}

Given these data, we gather performance metrics associated with normalization, forecasting, and inference approaches. For normalization, we consider working with the original, untransformed data, the DESeq2 size-factor normalized data \citep{Love2014ModeratedEO}, and the size-factor normalized data followed by an $\text{asinh}$ transformation \citep{Callahan2016BioconductorWF, Jeganathan2021ASP}. The latter two transformations account for potentially different library sizes across simulated samples and the fact that negative binomial data can be heavily skewed. For forecasting, we apply MDSINE2, fido with kernel parameters $\rho = \sigma = 0.5$ and $\rho = \sigma = 1$, and mbtransfer with $Q = P = 2$ and $Q = P = 4$. As mentioned above, MDSINE2 is an alternative to the gLV model designed to account for microbiome community dynamics \citep{Gibson2021IntrinsicIO}, and fido is a logistic-normal Gaussian Process model  \citep{silverman2022} (see Supplementary Section \ref{subsec:reference_methods} for details). All the models were provided with host and perturbation-related covariates. For MDSINE2, we forecast by integrating the learned dynamics over future timepoints, setting the initial conditions equal to the current test sample's microbiome community profile\footnote{We use \texttt{md2.integrate} as discussed in this MDSINE2 \href{https://github.com/gerberlab/MDSINE2_Paper/blob/ca884cd2f846560c075a4441414b90c5baae18d7/scripts/synthetic/helpers/evaluate.py}{documentation}.}. We consider $3 \text{ normalizations } \times 5 \text{ models} \times 27$ datasets, but exclude MDSINE2 on runs with 400 taxa due to consistently long computation times. This results in 378 simulation configurations.

We compare the cross-validated forecasting performance of the models across the simulation settings. We divide the 50 simulated subjects into $K = 4$ folds. For each iteration $k$, models are trained with the $\*y_{t}^{(i)}, \*w_{t}^{(i)}$ from all subjects except those in the holdout fold. On holdout folds, we reveal all timepoints up to the first intervention $t^{\ast}$ in the currently held-out subject. The trained models then forecast the community profiles up to a time horizon of $H = 5$. We provide access to intermediate interventions $\*w_{t + h}^{(i)}$, but not community compositions $\*y^{(i)}_{t + h}$ for $h > t^{\ast}$. For each iteration $k$, we compute the mean absolute error across lags, holdout subjects, and taxa:
\begin{align*}
MAE_{k} &= \frac{1}{JH}\frac{1}{\left|\mathcal{D}_{-k}\right|} \sum_{j = 1}^{J} \sum_{h = 1}^{H} \sum_{i \in \mathcal{D}_{-k}}\left|y^{(i)}_{j(t^{\ast} + h)} - \hat{f}_{j}^{+h}\left(y^{(i)}_{j\left(t^{\ast} - P + h\right):\left(t^\ast + h - 1\right)}, \*w^{(i)}_{\left(t^{\ast} - Q + h + 1\right):\left(t^{\ast} + h\right)}, \*z^{(i)}\right)\right|.
\end{align*}
We also evaluate inferential quality using false discovery proportions and power. Specifically, for instantaneous effects at lag $h = 0$, we compute the false discovery proportion and power as:

\begin{minipage}{.4\linewidth}
\begin{align*}
	\text{FDP}\left(0\right) = \frac{\left|J_{0} \cap \hat{J}\left(0\right)\right|}{\left|\hat{J}\left(0\right)\right|},
	\end{align*}
\end{minipage}
\begin{minipage}{0.4\linewidth}
	\begin{align*}
	\text{Power}\left(0\right) = \frac{\left|J_{1} \cap \hat{J}\left(0\right)\right|}{\left|J_{1}\right|},	
	\end{align*}
\end{minipage}

where $\hat{J}\left(0\right)$ are the taxa flagged as having immediate intervention effects and $J_{1}\left(0\right)$ are the rows of $B_{0}$ with at least one nonnull effect: $\cup_{d} \{j : B_{0,jd} \neq 0\}$. For delayed effects, we must account both for taxa with nonzero entries of $B_{q}$ for $q > 0$ and also those taxa that, though not directly influenced through $B_{q}$, are indirectly shifted by autoregressive links $A_{p}$ with taxa that are affected by the intervention. To this end, we recursively define:
\begin{align*}
	J_{1}\left(h\right) = &\left\{j : \text{row }j \text{ of } \prod_{p = 1}^{h} A_{p} \mathbf{1}_{J_{1}\left(h - p\right)} \text{ has at least one nonzero element}\right\} \bigcup \\ &\Bigl\{j : B_{h,jd} \neq 0 \text{ for some }d\Bigr\},
\end{align*}
where $\mathbf{1}_{J_{1}\left(h - p\right)}\in \{0, 1\}^{J}$ is an indicator over taxa that are nonnull at lag $h - p$. $J_{0}\left(h\right)$ is defined as the complement of $J_{1}\left(h\right)$. The mirror-selected taxa at delay $h$ are denoted $\hat{J}\left(h\right)$, and they can be compared with $J_{1}\left(h\right)$ and $J_{0}\left(h\right)$ to define $\FDP\left(h\right)$ and $\text{Power}\left(h\right)$.

\subsection{Results}
\label{sec:results}

\begin{figure}[!p]
\centering\includegraphics[width=\textwidth]{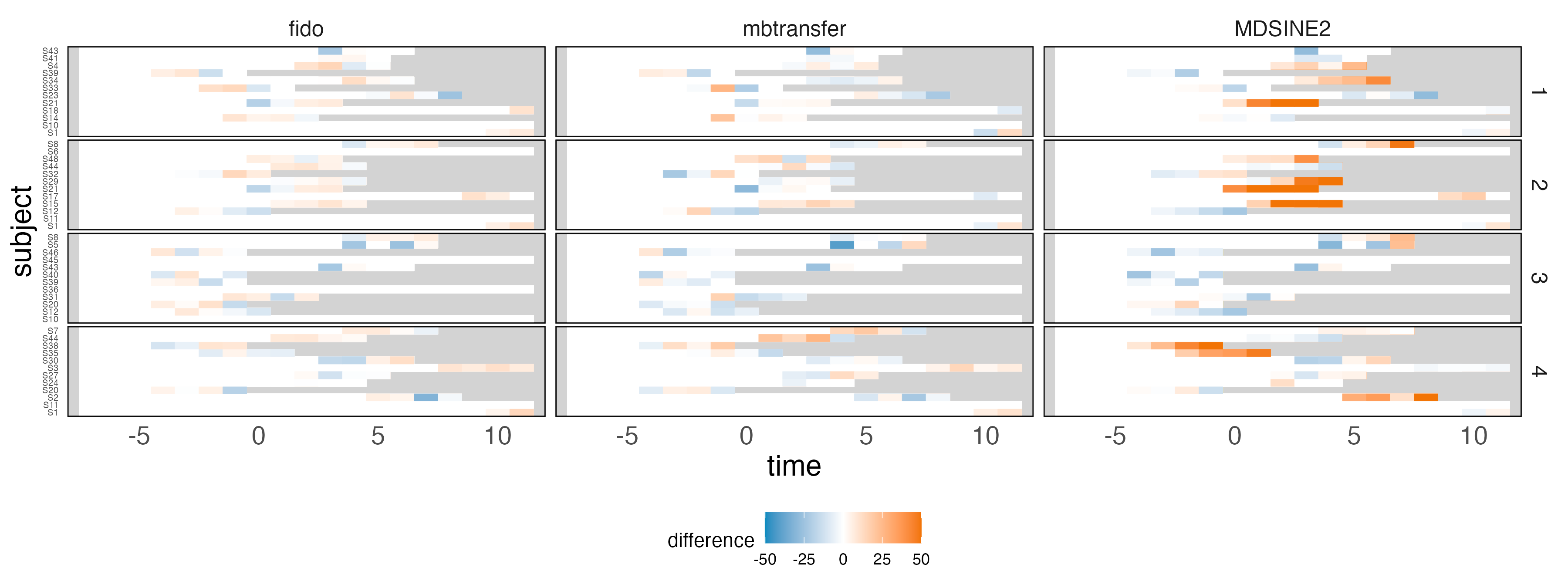}
	\caption{A comparison of forecasting residuals across four folds (rows) in one simulation run suggests that forward integrating the MDSINE2 model can lead to exponential increases in forecasting errors.}
	\label{fig:differenced_data}
\end{figure}

\begin{figure}[!p]
\centering\includegraphics[width=\textwidth]{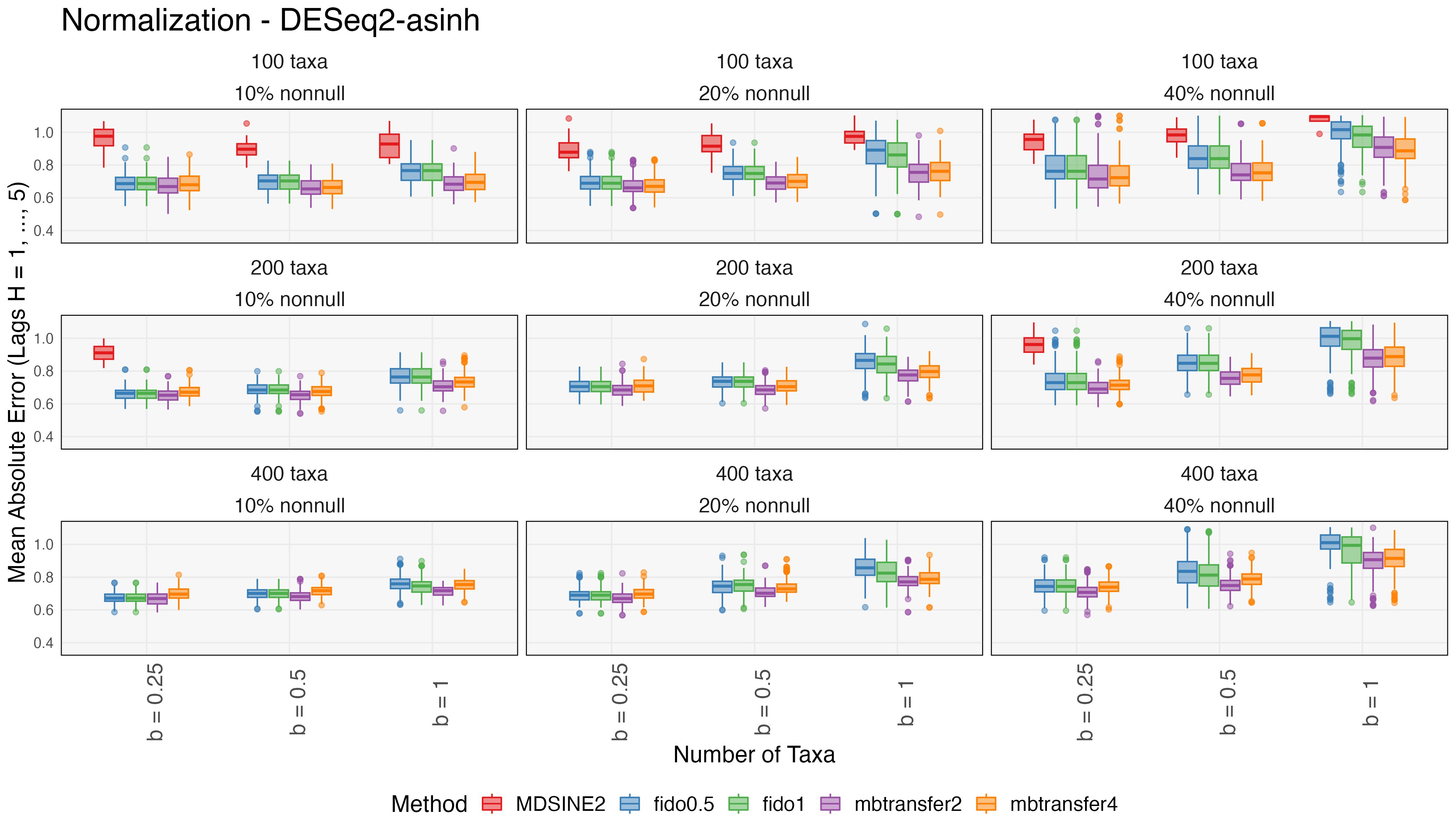}
\caption{Forecasting errors across simulation settings.
	The $y$-axis corresponds to the average of $MAE_{k}$ across folds. Within each panel, the signal strength $b$ increases from left to right. Column panels give the proportion of taxa affected by the intervention, and rows have different numbers of taxa. Errors beyond 3$\times$ the interquartile range of those from fido and mbtransfer have been omitted from the view, excluding some outliers from MDSINE2. Runs that did not complete within 72 hours are not included -- this explains the missing boxplot for MDSINE2 in the 200 taxa, 20\% nonnull panel. The fido package is comparable to mbtransfer when the intervention strength is weak but deteriorates when the intervention is strong.} 
\label{fig:forecasting-sim-DESeq2-asinh}
\end{figure}

Figure \ref{fig:forecasting-sim-DESeq2-asinh} summarizes cross-validated forecasting performance on DESeq2-asinh transformed data. We also discuss alternative transformations below. Error rates increase with the proportion of nonnull taxa $1 - \pi_{0}$ and signal strength $b$. This increase is likely a consequence of the high variance shifts in $\*y_{t}$ during interventions for these settings. MDSINE2's performance is consistently worse than either fido or mbtransfer's. Figure \ref{fig:differenced_data} sheds light on this. It shows prediction error for individual holdout subjects in one of the simulation settings; residual error in other simulation settings is qualitatively similar. We average errors across all taxa and truncate those with a magnitude greater than 50. This figure shows that minor errors in MDSINE2's initial forecast become amplified at larger horizons. This behavior is not universal, but its effects on the subset of subjects where it does appear are strong enough to explain MDSINE2's deterioration in our simulation setup. In retrospect, such behavior is unsurprising -- MDSINE2 can only refer to one step in the past, and it must have either exponential growth or decay until the community reaches its carrying capacity. Though this behavior does not affect inferences for taxon-perturbation relationships, which are the main focus of \citep{Gibson2021IntrinsicIO}, it can limit the usefulness of the forecasts needed to simulate hypothetical trajectories. In contrast, mbtransfer and fido can refer to historical windows, supporting more realistic intervention analysis: The second day of a microbiome intervention does not necessarily have the same consequences as the first day.

When the intervention effect has a smaller magnitude or is limited to fewer taxa, fido and mbtransfer perform comparably. In other cases, mbtransfer is more accurate. We interpret this by noting that, despite its ability to incorporate interventions as covariates, fido's Gaussian Process assumption enforces smoothness in the predicted values. This smoothness prevents the model from capturing the sharp changes in abundance within these simulation settings.
Since the simulation's true autoregressive dynamics have $P = Q = 3$, the fitted mbtransfer models are misspecified. Interestingly, the $P = Q = 2$ model slightly outperforms the $P = Q = 4$ model. In this context, the reduction in variance from having a slightly less flexible model outweighs the reduction in bias from modeling larger lags. 

Analogous results for alternative transformations are available in Supplementary Figures \ref{fig:forecasting-sim-none} and \ref{fig:forecasting-sim-DESeq2}. When the data are not asinh transformed, the mbtransfer model performs worse than either MDSINE2 or fido. This reversal is consistent with the use of a squared-error loss in the underlying gradient boosting model, which is not adapted to count data. fido should be preferred if data must be modeled on the original scale. However, we note that transformations are often well-justified in microbiome analysis, and an increasing number of formal methods implement them \citep{McKnight2018MethodsFN, Chen2018GMPRAR, Jiang2021mbImputeAA}. Finally, Supplementary Figure \ref{fig:computation-sim} gives the average computation time across folds. MDSINE2 is slower than either fido or mbtransfer. fido and mbtransfer have comparable computation times except when using DESeq2-asinh transformed data. In this setting, fido is noticeably faster, but mbtransfer provides better forecasts.

%Figs 3 \& 4

\begin{figure}[!p]
\centering\includegraphics[width=\textwidth]{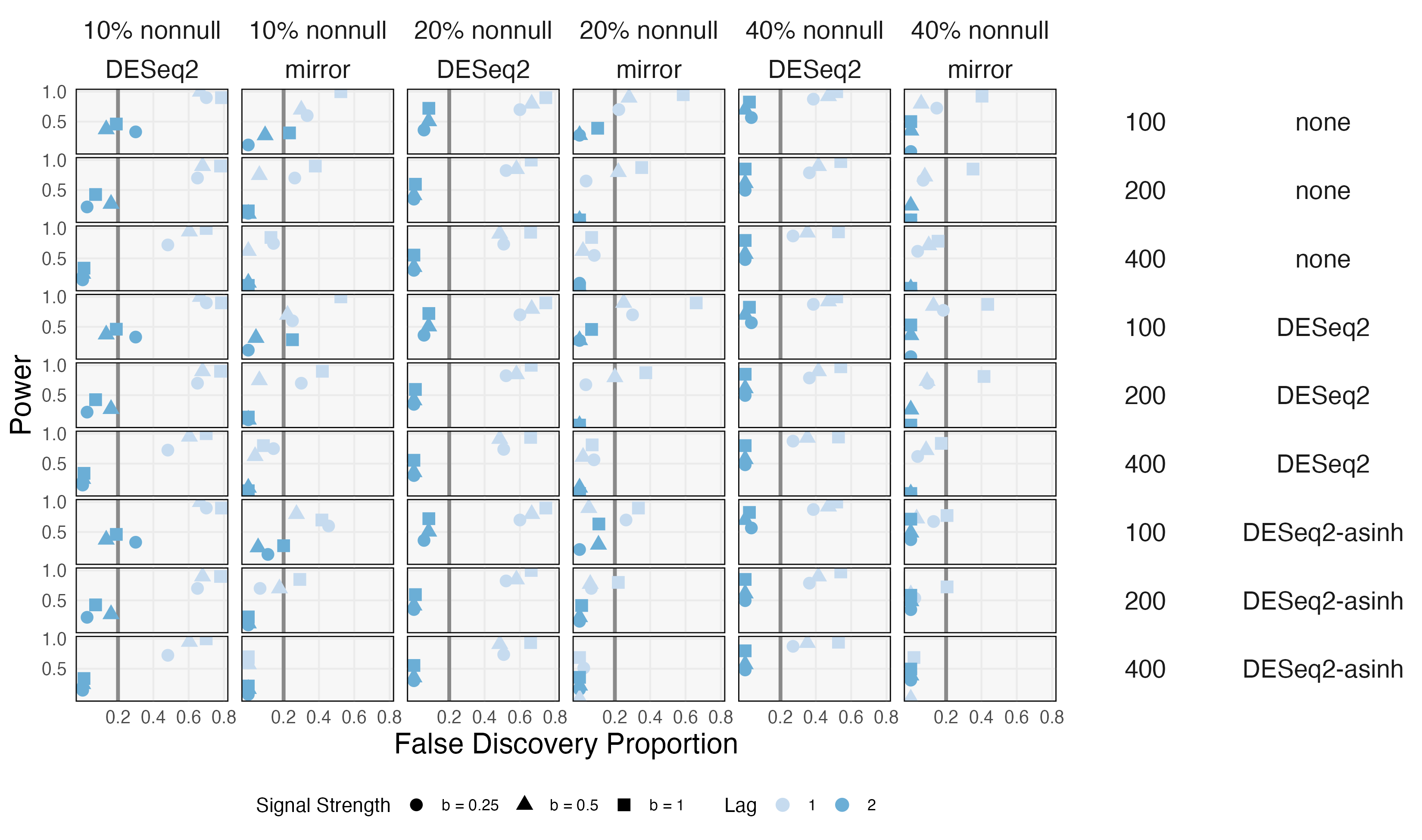}
	\caption{Inferential performance in the simulation experiment. Rows index different normalization methods and the total number of taxa. The target FDR in each case has been set to $0.2$. Columns have varying proportions of nonnull hypotheses and compare DESeq2-based inferences with those from mirror statistics. DESeq2 does not provide FDR control for lag one effects in any simulation context. mbtransfer's mirror algorithm controls the FDR when using the DESeq2-asinh transformation data and when the number of taxa is sufficiently large.}
	\label{fig:inference-sim-error}
\end{figure}

Figure \ref{fig:inference-sim-error} summarizes inferential performance. When considering longer time horizons, all methods have improved FDR control because more taxa become truly nonnull as instantaneous effects propagate across the community. However, DESeq2 never appears to control the FDR at the prespecified level of $q = 0.2$. Though DESeq2 assumes a negative binomial generative mechanism across taxa, it treats samples as independent. This misspecification likely contributes to the poor FDR control seen in this simulation. For larger $b$, the mirror statistics may appear conservative, with many $\widehat{\FDP} \ll 0.2$. These settings often correspond to high power, though, and the signal may have simply become easy to detect. Mirrors control the FDR when the DESeq2-asinh transformation is applied, and the number of taxa is large. This is expected, because the DESeq2-asinh transformation improves forecasts, and the need for a large number of taxa is consistent with Proposition 3.3 of \citet{Dai2020FalseDR}, which demonstrates FDR control only asymptotically. We attribute the strong performance of mirror statistics to the fact that its false discovery rate control is adapted to the current dataset of interest rather than a previously defined probabilistic model. 

\section{Data Analysis}
\label{sec:data_analysis}
We next illustrate mbtransfer using three microbiome datasets. The data are drawn from two human and one animal microbiome studies. They include an experimentally-defined intervention, one that arises in the natural progression of a prospective study, and a shift in continuous ecosystem parameters. The studies also vary in their taxonomic richness, number of subjects, and total number of timepoints. Despite the various domains and intervention types considered, each study focuses on how environmental change remodels the microbiome.

\subsection{Diet and the gut microbiome}
\label{subsec:diet_gut_microbiome}

\citet{David2013DietRA} investigated the sensitivity of the human gut microbiome to brief diet interventions. To this end, they recruited 20 participants and randomly assigned them to either ``plant'' or ``animal'' interventions. Subjects in the two groups were required to maintain a plant- or animal-based diet during a five-day intervention window. Samples were collected for two weeks surrounding the intervention, typically at a daily frequency. Ultimately, 8 - 15 samples were collected for each participant since some timepoints were never successfully sampled. We linearly interpolate timepoints onto an even, daily sampling grid, motivated by the cubic spline interpolation adopted by \citet{RuizPerez2019DynamicBN}. Regularly spaced timepoints are a fundamental limitation of discrete, autoregressive models. Initially, the data contained 17310 taxa. We filter to those present in at least 40\% of the samples, reducing the number of taxa to 191 -- a drastic reduction, but one consistent with distinguishing a ``core'' microbiome for more focused analysis \citep{Shade2012BeyondTV, Neu2021DefiningAQ}.

\begin{figure}[!p]
\centering\includegraphics[width=\textwidth]{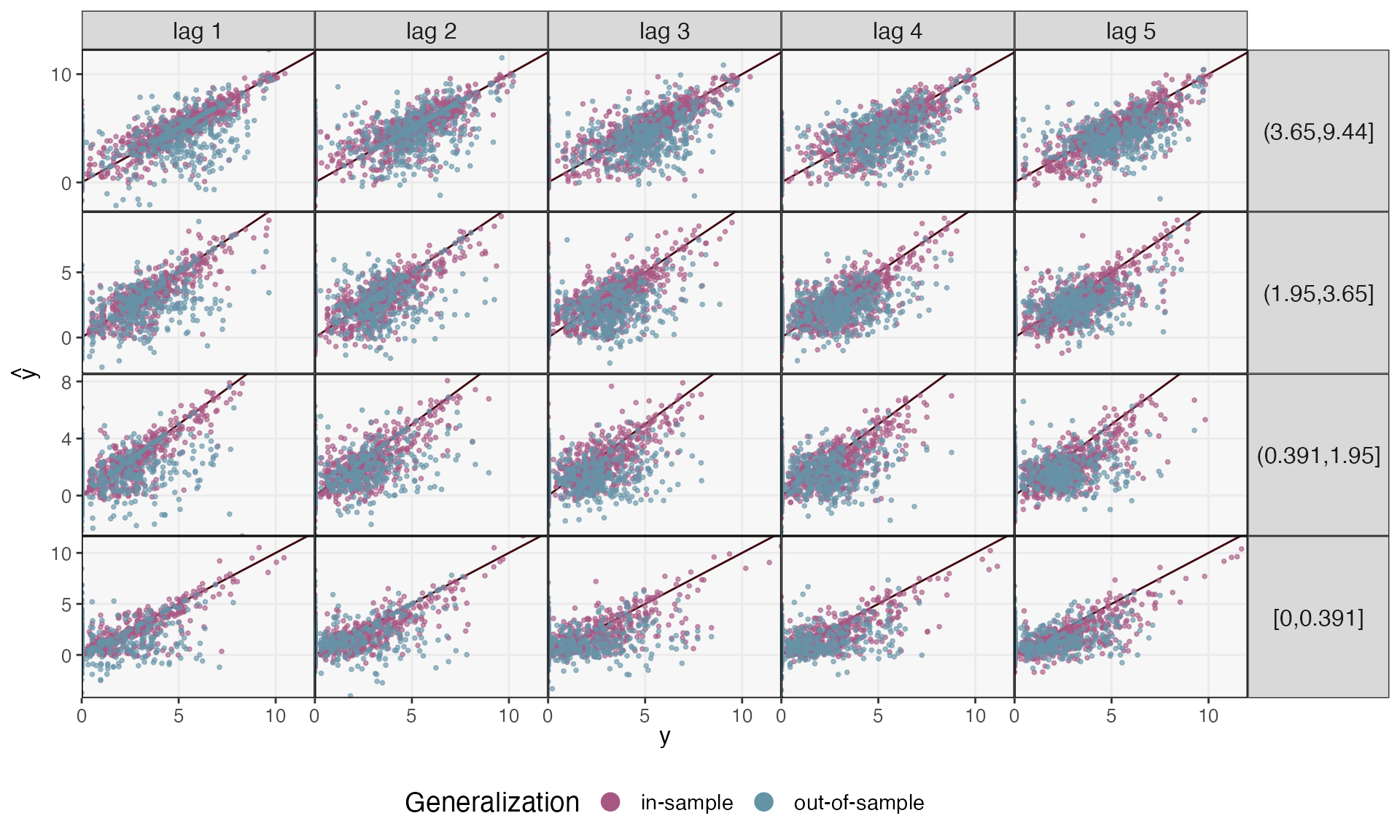}
	\caption{Forecasting error for an mbtransfer model applied to the diet intervention dataset of Section \ref{subsec:diet_gut_microbiome}. The $y$-axis is faceted by quantiles of abundance and the $x$-axis is faceted by time horizon $h$. In-sample error refers to errors made on new timepoints for individuals who appeared in the training data, while out-of-sample predictions are made on individuals who did not appear during training. Performance is strongest in shorter time horizons and for more abundant taxa.}
	\label{fig:diet_forecast_error}
\end{figure}

We fit mbtransfer with $P = Q = 2$ and $\*w^{(i)}_{t} = \left(\indic{t \in \text{Animal Shift}}, \indic{t \in \text{Plant Shift}}\right) \in \left[0, 1\right]^{2}$, setting two intervention series\footnote{It is possible for these series to lie between 0 and 1 because some interpolated timepoints lie in the transitions between active and inactive periods -- see Figure \ref{fig:opening}.} and omitting any host features $z^{(i)}$. Figure \ref{fig:diet_forecast_error} shows in- and out-of-sample forecasts. Forecasting performance deteriorates out of sample, highlighting the between-participant heterogeneity in this study. This challenge is most pronounced within the lowest quantile of abundance. Nonetheless, out-of-sample forecasts are still clearly correlated with the truth. In both the in and out-of-sample contexts, forecasts on shorter time horizons are more accurate. In addition, performance seems better in the more highly abundant taxa. The gradient boosting model's least squares training objective likely deteriorates in sparse data.

\begin{figure}[!p]
\centering \includegraphics[width=\textwidth]{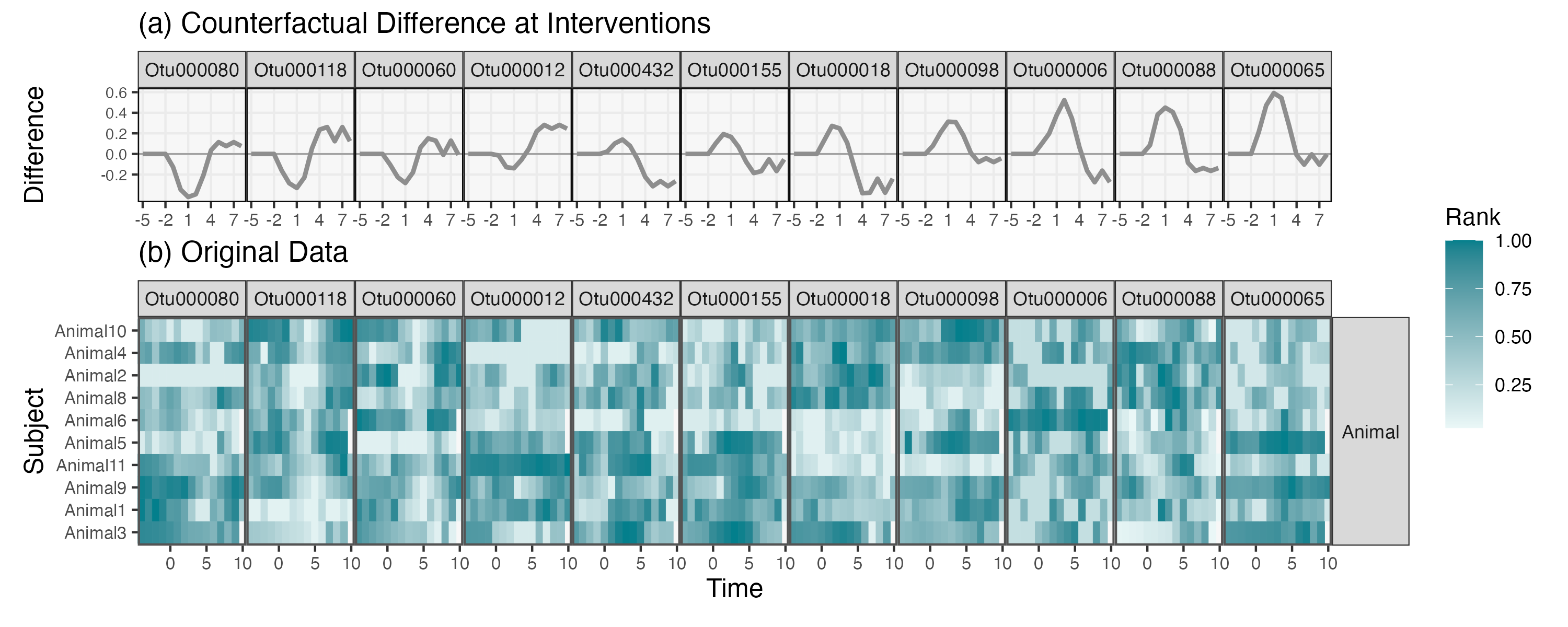}
    \caption{(a) Counterfactual difference in simulated trajectories for a subset\protect\footnotemark  of the selected taxa in the diet data in Section \ref{subsec:diet_gut_microbiome}. (b) Subject-level data from a subset of taxa appearing in (a). Each heatmap row is a subject, and each column is a timepoint. These data are consistent with the interpretations from the counterfactual simulation. For example, OTU000006 has more transient increases in abundance (e.g., Animal1, and Animal6) while OTU000065 has more prolonged departures (e.g., Animal3 and Animal 9).}
    \label{fig:diet-combined}
\end{figure}

We compute mirror statistics for time lags $h = 1, \dots, 4$ to evaluate the effect of a four day shift to an animal diet. Supplementary Figure \ref{fig:diet_mirror_statistics} shows the associated mirror statistic distributions.
The increasing magnitude across lags for some taxa suggests that the diet intervention effects are not instantaneous but build up over consecutive treatment days. To support interpretation, Figure \ref{fig:diet-combined}a shows the median difference between counterfactual trajectories for a subset of significant taxa. The taxa were chosen by applying principal component analysis to the simulated trajectory differences, projecting onto the first component, and selecting every sixth taxon according to that ordering.  
Some taxa (e.g., OTU000006) have more immediate but transient effects, while others (e.g., OTU000065) have more gradual but sustained changes. Further, in several taxa (e.g., OTU000118, OTU000012), a long-run increase follows an initial decrease, which is corroborated by the associated subject-level data. Within taxa, we found that the first and third quartiles of the counterfactual differences across subjects tended to agree. This suggests that the model has not learned interactions between the intervention effects and past composition -- the effect of the diet intervention may be uniform across various initial community states. The main benefit of a transfer function modeling approach is the model's capacity to learn different shapes of counterfactual trajectories while still controlling a precise notion of FDR.

For comparison, the original, interpolated data for a subset of taxa is shown in Figure \ref{fig:diet-combined}b. These views are consistent with the counterfactual trajectories, but they are obfuscated by the higher degree of sampling noise and require more space. Our results are in line with \citet{David2013DietRA}, but we can clearly describe ecosystem dynamics by modeling temporal dependence between the diet intervention and microbiome community profiles.

\subsection{Birth and the vaginal microbiome}
\label{subsec:postpartum}

\begin{figure}[!p]
\centering \includegraphics[width=\textwidth]{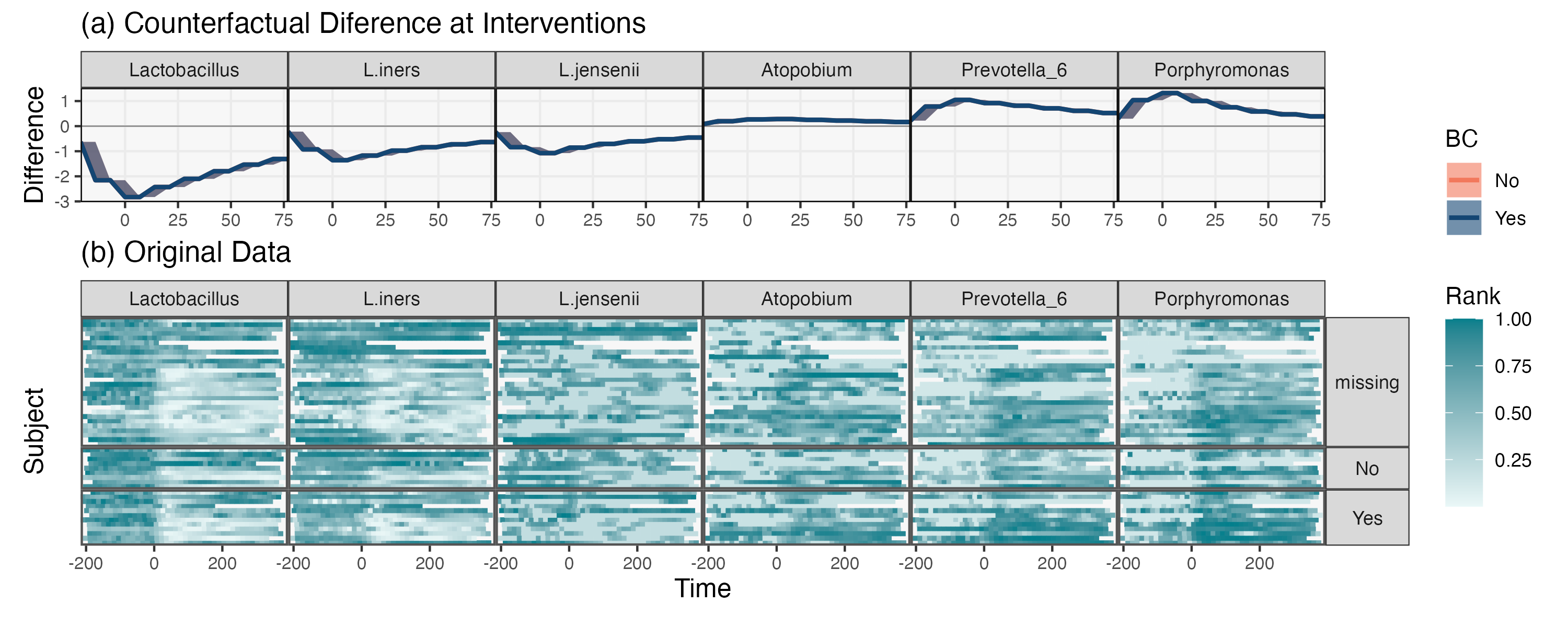}
    \caption{(a) Counterfactual differences for a subset of selected taxa from the re-analysis of \citep{Costello2022LongitudinalDO}. Counterfactual differences are computed for each subject in the data, and bands represent the first and third quartiles of differences across subjects. Since the bands for birth control reinitiation overlap, we conclude that the model does not learn the interaction effects between the intervention and contraception use. (b) The corresponding subject-level data. Rows are grouped according to the birth control reinitiation survey response.}
    \label{fig:postpartum-combined}
\end{figure}

We next re-analyze data from \citet{Costello2022LongitudinalDO}, which studied how birth influences the composition of the mother's vaginal microbiome. Supplementary Figure \ref{fig:postpartum_forecast_error} shows in and out-of-sample forecasting accuracy. Compared to Figure \ref{fig:diet_forecast_error} in the previous analysis, in and out-of-sample performances are more comparable, reflecting the larger sample size of this study. The derived mirror statistics are shown in Supplementary Figure \ref{fig:postpartum_mirror_statistics}. Compared to the diet intervention, more of the ecosystem is shifted by the birth intervention. We generate four counterfactual trajectories for all subjects to understand how birth influences individual taxa and whether any effects are modulated by contraception use. Specifically, we compute $\hat{f}^{+h}\left(\*y_{(t^{\ast} - P - 1):(t^{\ast} - 1)}), \tilde{\*w}_{(t^{\ast} - Q):t^{\ast}}, \tilde{z}\right)$ for $\tilde{\*w}_{(t^{\ast} - Q):t^{\ast})} \in \{\*1_{Q}, \*0_{Q}\}$ representing presence or absence of the birth event and $\tilde{z} \in \{0, 1\}$ denoting re-initiation of contraceptive use following birth. Figure \ref{fig:postpartum-combined}a suggests the absence of an interaction with contraceptive use. This may be a consequence of the fact that 57\% of subjects were missing any data on contraceptive use -- though the examples discussed in \cite{Costello2022LongitudinalDO} consider plausible mechanisms for how contraception can influence the postpartum microbiome, our model does not detect a generalizable enough association to learn the interaction.

\footnotetext{Specifically, we include panels for every sixth selected taxon after sorting according to the first dimension of the PCA taken on simulated trajectories.}

Like in the diet intervention, we can distinguish between response trajectories. Members of genus \textit{Lactobacillus} are clearly depleted, while other taxa appear to take advantage of the postpartum environment. For example, \textit{Porphyromonas} appears briefly during the same window that the \textit{Lactobacilli} disappear. Figure \ref{fig:postpartum-combined}b compares these trajectories with real data. As before, we see that the learned trajectories denoise the original data, and consistent with the lack of interaction, we do not observe obvious, systematic associations between postpartum community trajectory and contraception use.

\subsection{pH and the aquaculture microbiome}
\label{subsec:aquaculture}

We next use mbtransfer in a problem with continuous intervention values. 
\citet{Yajima2022CoreSA} studied the taxonomic composition of the eel aquaculture microbiome, collecting water samples every 24 hours for 128 days from five aquaculture tanks. We can view the tank's pH and eel activity scores as continuous inputs $\*w_{t}$ to a transfer function model. Based on the five tanks' longitudinal data, \cite{Yajima2022CoreSA} concluded that the microbiome composition changes over time and is related to various environmental factors. Moreover, there was a substantial shift in pH in three of the tanks, and we analyze how this shift influenced community composition.

\begin{figure}[!p]
    \centering
    \includegraphics[width=\textwidth]{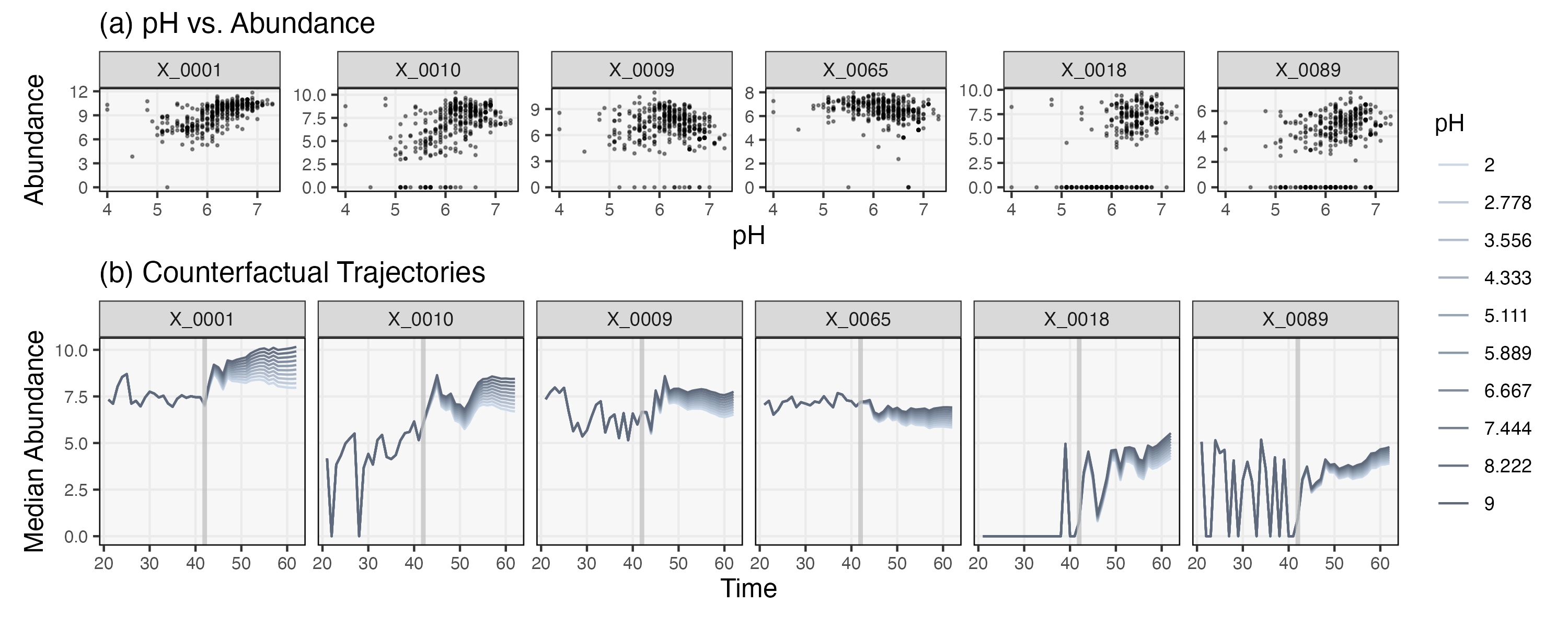}
    \caption{(a) Associations between pH and abundance for a subset of taxa that are selected by the mirror algorithm. Taxa have been sorted in decreasing order according to the magnitude of their associated mirror statistics. (b) Counterfactual trajectories under pH shifts. A sustained pH shift is imagined starting at day 42. Different values of pH tend to influence the magnitude, but not the shape, of the forecast trajectories}
    \label{fig:aqua-combined}
\end{figure}

After preprocessing the 16S data, we have 345 samples and 128 taxa. We interpolate the sampling times to fill in some missing days. Then, we fit a mbtransfer with $P =4$ and $Q =4$. We simulate counterfactual pH series that are constant for ten timepoints, with values varying between pH $=2$ and $=9$. Figure \ref{fig:aqua-combined} shows scatterplots of pH vs. abundance for a subset of taxa that were found to be significant when contrasting the two extremes, pH = $2$ and $9$. Moreover, we can also simulate taxa trajectories for each counterfactual pH series (Figure \ref{fig:aqua-combined}b). The effects seem additive, with counterfactual abundances smoothly varying as a function of pH. The taxa with the clearest associations (e.g., X\_0001, X\_0010) appear to have larger variation in their simulated trajectories, which agrees with their being more sensitive to changes in pH.

\section{Software}

We created the mbtransfer R package for analyzing interventions using transfer function modeling. This package provides various functionalities, including the creation of an S4 object called \texttt{ts\_inter}, handling intervention windows, conducting counterfactual simulations, and performing inference based on mirror statistics. The block below creates an example \texttt{ts\_inter} container that unifies intervention time series and microbiome profiles,
\begin{minted}{R}
ts <- as.matrix(reads) |>
  ts_from_dfs(interventions, samples, subject)
\end{minted}

We can generalize the time delay for any number of input covariates. In addition, if there are nonuniform sampling time points, we can use \texttt{interpolate()} to get different resolutions of sampling:

\begin{minted}{R}
ts <- ts |>
  interpolate(delta = 1, method = "linear")
\end{minted}

After setting up the data, we can train a transfer function-based boosting model. The model assumes that interventions at different time lags impact microbial communities within a specified range of time lags. For instance, we can train the model with $P=3$ and $Q=3$, where $P$ represents the maximum time lags affecting the communities and $Q$ denotes the range of intervention time lags,
\begin{minted}{R}
fit <- mbtransfer(ts, P = 3, Q = 3)
\end{minted}

Once the model is trained, we can forecast future community composition. \texttt{predict()} will fill in any time points that are present in the \texttt{intervention} slot, which allows forecasts across different time horizons. For example, the block below fills in predicted compositions from the ninth time point on,
\begin{minted}{R}
ts_preds <- list()
ts_preds <- predict(fit, subset_values(ts, 1:8))
\end{minted}

Following model evaluation, we can identify taxa with instantaneous or delayed intervention effects. This can be achieved by simulating counterfactual alternatives and employing partial dependence mirror statistics to control the false discovery rate. We can obtain a list of intervention series by specifying the characteristics of the hypothetical interventions (e.g., step interventions starting at specific time points and with defined lengths). 

\begin{minted}{R}
ws <- steps(c("D1" = TRUE), starts = 1, lengths = 2:4, L = 4)
\end{minted}

Using these series and the trained model, we can select taxa with significant effects based on a specified false discovery rate.

\begin{minted}{R}
staxa <- select_taxa(ts, ws[[1]], ws[[2]], ~ mbtransfer(., 3, 3))
\end{minted}

As transfer function-based boosting models can incorporate various covariates, this framework enables the identification of factors that influence shifts in each taxon.

\section{Discussion}

mbtransfer adapts transfer function models to the dynamic microbiome context. The approach is flexible and interpretable, enabling intervention analysis without assuming a restrictive functional form and supporting the simulation of counterfactual trajectories. We have complemented our modeling approach with a formal inferential mechanism, leveraging recent advances in selective inference. A simulation study illuminated our method's properties across data-generating settings, and our data analysis highlighted its practical application in three contrasting microbiome studies.

We anticipate several directions for further study. First, we have focused attention on developing mirror statistics for detecting temporal effects, but the same strategy could be generalized to support inference for inter-species relationships and host-microbiome interactions. Indeed, the construction of mirror statistics via partial dependence profiles depends only on having access to a simulator $f$ that can generate hypothetical responses. This simulator could be used to contrast profiles with alternative initial states $\*y_{t}$ or host features $\*z$. Similarly, the procedure could clarify interactions between several concurrent interventions or scales \citep{Fukuyama2021MultiscaleAO, Sankaran2022GenerativeMA}. Second, it would be valuable to develop a transfer function model that learns the entire distribution of responses $p\left(\*y_{t} \vert \*y_{(t - P - 1):(t - 1)}, \*w_{(t - Q + 1):t}\right)$, rather than simply the mean. Such a probabilistic analog would allow us to quantify uncertainty in intervention effects. Characterizing the uncertainty in intervention effects is especially valuable in the design of potential probiotics -- an intervention with moderate but consistent effects may be preferable to one with strong but erratic ones \citep{Thompson2022IntegratingAT, Fannjiang2022ConformalPU, Jeganathan2018TheBB}. Finally, an extension to continuous time autoregressive processes would allow us to model irregular sampling frequencies, removing the need for the interpolation steps performed above.

We have synthesized a variety of statistical concepts to address a recurring microbiome data analysis challenge: How can we quantify the influence of environmental shifts on a microbial ecosystem? The transfer function perspective has guided our intervention analysis, and we linked the resulting nonlinear models with a modern computational inference technique based on mirror statistics. This facilitates the stability and attribution analysis critical for forming scientific conclusions \citep{Efron2020PredictionEA, Yu2018ThreePO}. As microbiome studies continue to investigate more and more nuanced questions about ecosystem dynamics, similarly formal simulation and inference methods will likely play an essential role.

%\pagebreak

\section{Supplementary Material}
\label{supp}

Supplementary material is available online at
% \href{http://biostatistics.oxfordjournals.org}%
% {http://biostatistics.oxfordjournals.org}.
\url{http://biostatistics.oxfordjournals.org}.

\section*{Acknowledgements}

Support for this research was provided by the University of Wisconsin - Madison Office of the Vice Chancellor for Research and Graduate Education with funding from the Wisconsin Alumni Research Foundation.

P. Jeganathan received funding from the Faculty of Science at McMaster University. 

{\it Conflict of Interest}: None declared.

\bibliographystyle{biorefs}
\bibliography{references.bib}

\pagebreak
%\documentclass{article}
%\usepackage{placeins}
%\usepackage{float}
%\usepackage{longtable}
%\usepackage{caption}
%\usepackage{xr-hyper}
%\usepackage{hyperref}
%\usepackage{natbib}

%\makeatletter
%\newcommand*{\addFileDependency}[1]{
%\typeout{(#1)}
%\@addtofilelist{#1}
%\IfFileExists{#1}{}{\typeout{No file #1.}}
%}\makeatother

%\newcommand*{\myexternaldocument}[1]{%
%\externaldocument{#1}%
%\addFileDependency{#1.tex}%
%\addFileDependency{#1.aux}%
%}

%\myexternaldocument{main}

%\begin{document}
\section*{Supplementary Materials}

\subsection{Reproducibility}
\label{subsec:reproducibility}

\subsubsection*{Simulation experiments}

\begin{figure}[H]
    \centering
    \includegraphics[width=\textwidth]{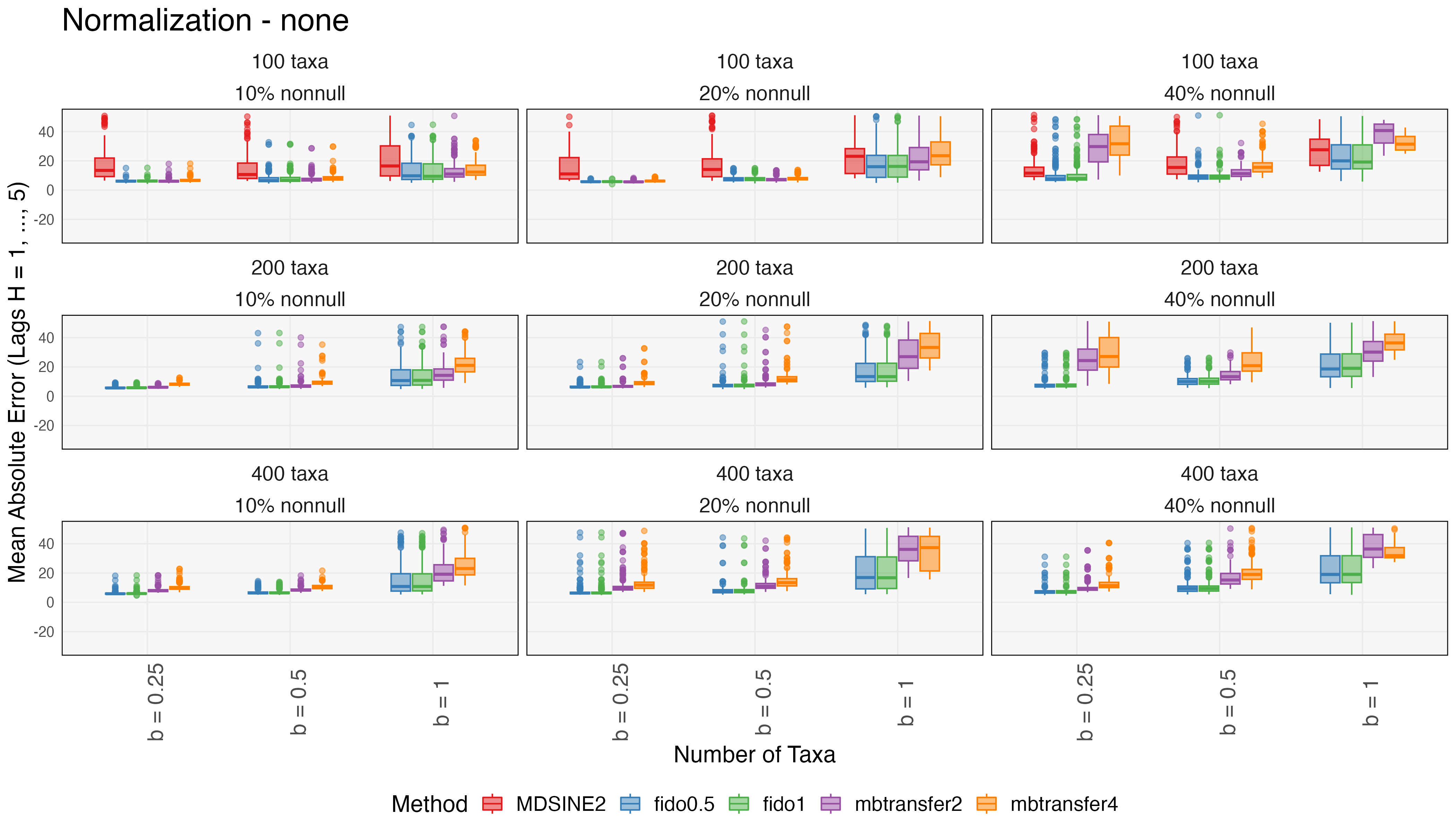}
    \caption{The analog of Figure 3 %\ref{fig:forecasting-sim-DESeq2-asinh} 
    when not using any normalization.}
    \label{fig:forecasting-sim-none}
\end{figure}

\begin{figure}
    \centering
    \includegraphics[width=\textwidth]{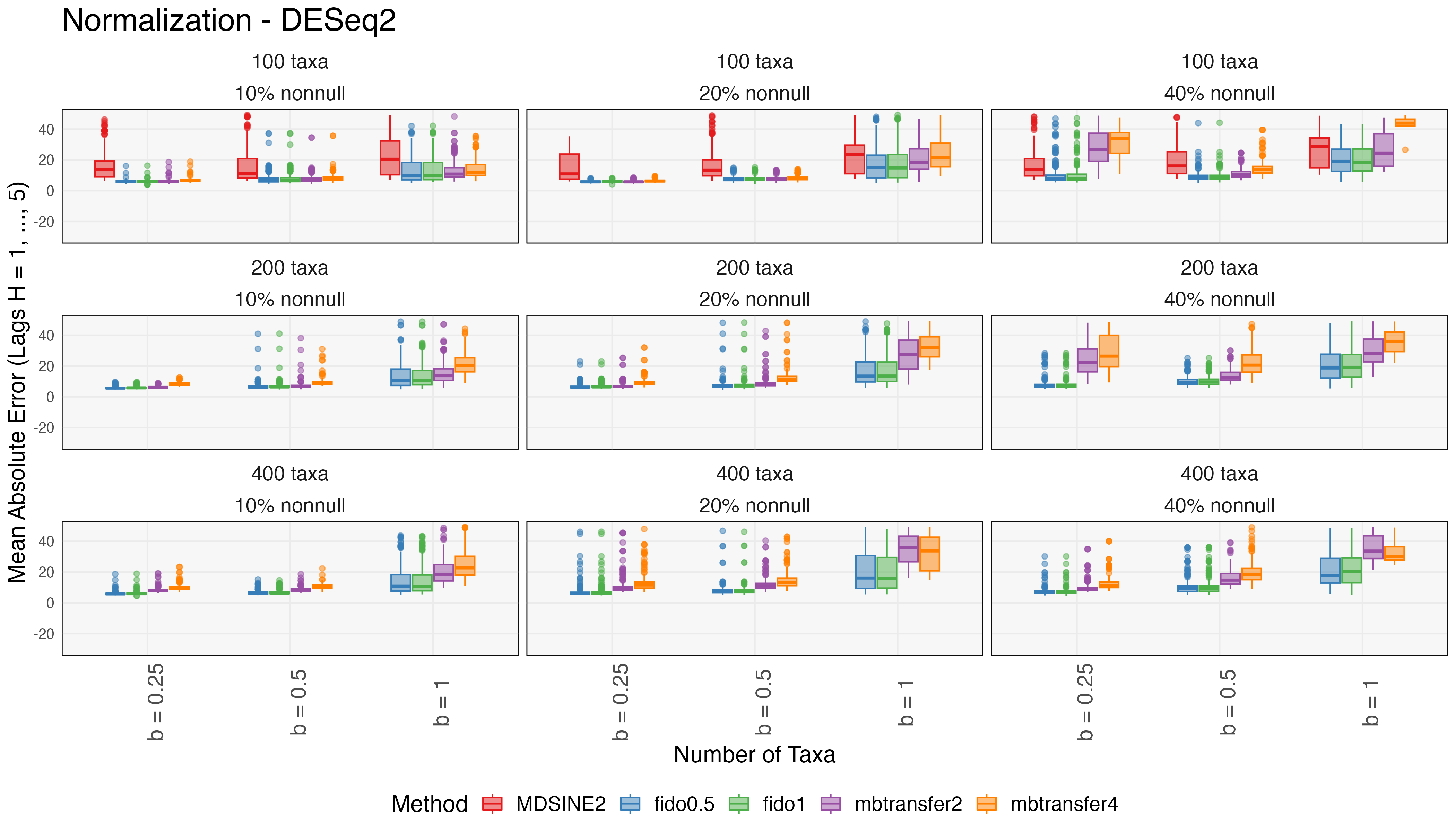}
    \caption{The analog of Figure 3 %\ref{fig:forecasting-sim-DESeq2-asinh} 
    when using DEseq2 size-factor normalization.}
    \label{fig:forecasting-sim-DESeq2}
\end{figure}

\begin{figure}
    \centering
    \includegraphics[width=\textwidth]{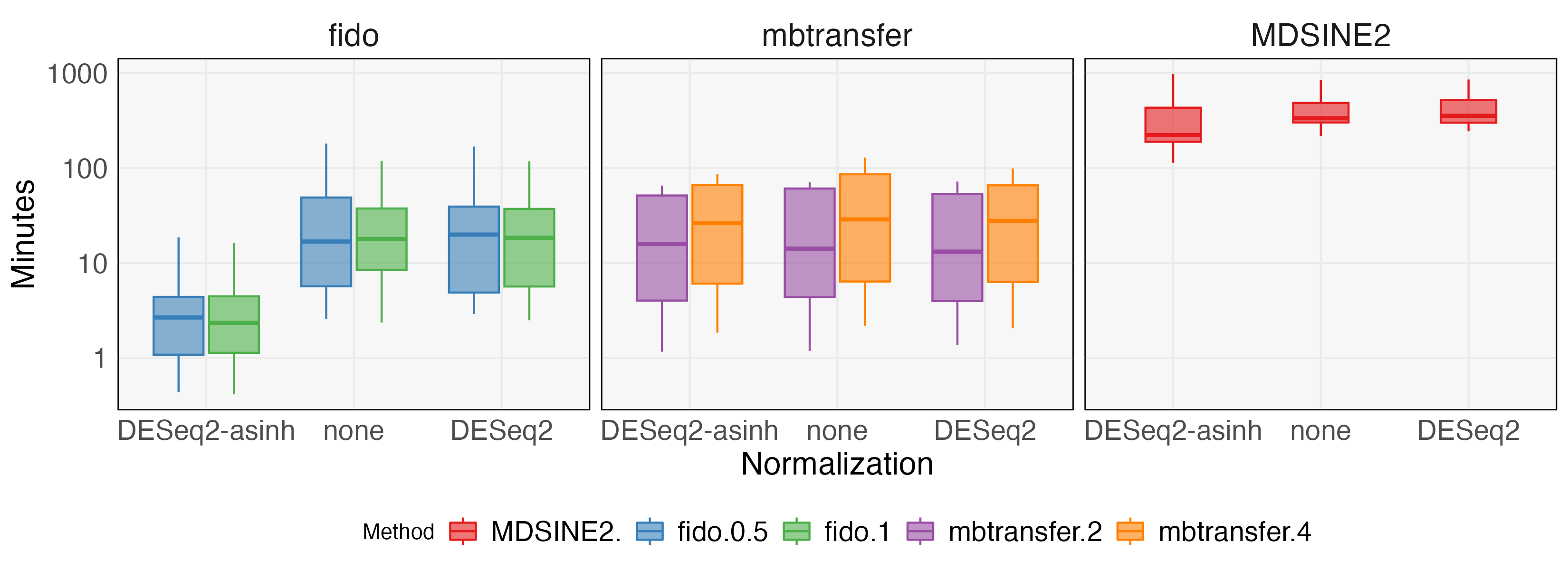}
    \caption{Computation times for methods considered in the simulation experiment. fido is fast on untransformed, count data, which is the context in which it was originally designed. mbtransfer is comparable to fido on transformed data. Both packages are an order of magnitude faster than MDSINE2.}
    \label{fig:computation-sim}
\end{figure}

\begin{figure}
	\includegraphics[width=0.9\textwidth]{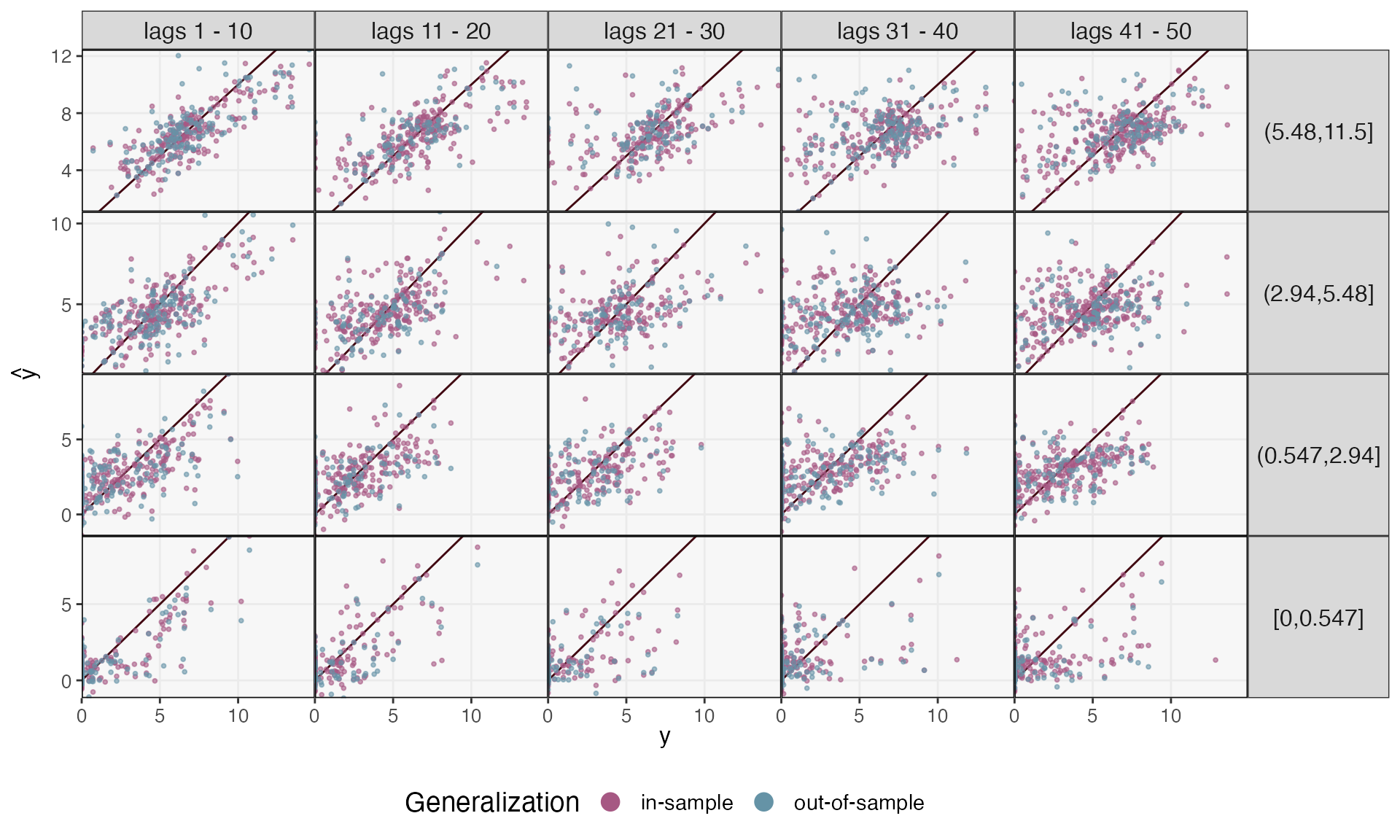}
	\caption{The analog of Figure 6 %\ref{fig:diet_forecast_error} 
 for the postpartum case study in Section 4.2. %\ref{subsec:postpartum}.
 Each point is one sample. In-sample error refers to errors from future timepoints of subjects observed in the training data. Out-of-sample errors are those on previously unobserved subjects. As before, errors predictions are most accurate on nearby time lags and more abundant taxa.}
  \label{fig:postpartum_forecast_error}
\end{figure}

\begin{itemize}
\item A Dockerfile that installs software used in the experiments is available at \href{https://go.wisc.edu/eovk4b}{https://go.wisc.edu/eovk4b}. The image can be pulled from DockerHub using \texttt{docker pull krisrs1128/mi:20230506}.
\item Simulation inputs have been saved at \href{https://go.wisc.edu/8ey754}{https://go.wisc.edu/8ey754}. They were generated using \href{https://go.wisc.edu/37y6ny}{https://go.wisc.edu/37y6ny}. This script also includes source code for Figure 2. %\ref{fig:simulation-hm}.
\item Each forecasting and inference simulation run corresponds to one \texttt{run\_id} of these Rmarkdown notebooks: \href{https://github.com/krisrs1128/microbiome_interventions/blob/main/scripts/forecasting_metrics.Rmd}{I}, \href{https://github.com/krisrs1128/microbiome_interventions/blob/main/scripts/inference_metrics.Rmd}{II}. Simulation outputs have been saved at \href{https://go.wisc.edu/3gc982}{https://go.wisc.edu/3gc982}.
\item Figures 3, 5, %\ref{fig:forecasting-sim-DESeq2-asinh}, \ref{fig:inference-sim-error}, %
and Supplementary Figures \ref{fig:forecasting-sim-none}, \ref{fig:forecasting-sim-DESeq2}, \ref{fig:computation-sim} were generated using this script \href{https://go.wisc.edu/87876s}{https://go.wisc.edu/87876s} applied to the previous outputs. Figure 4 %\ref{fig:differenced_data} 
was generated using \href{https://go.wisc.edu/l1n79m}{https://go.wisc.edu/l1n79m}.
\end{itemize}

\subsubsection*{Case studies}

\begin{itemize}
	\item The case studies appear as vignettes in our accompanying R package (\href{https://go.wisc.edu/crj6k6}{https://go.wisc.edu/crj6k6}, \href{https://krisrs1128.github.io/mbtransfer/articles/aqua.html}{I}, \href{https://krisrs1128.github.io/mbtransfer/articles/diet.html}{II}, \href{https://krisrs1128.github.io/mbtransfer/articles/postpartum.html}{III}). They can also be rerun without installing the package by visiting this binder notebook: \href{https://go.wisc.edu/emxv33}{https://go.wisc.edu/emxv33}.
	\item Processed versions of the data used in all case studies can be found on figshare: \href{https://go.wisc.edu/7ig8q8}{https://go.wisc.edu/7ig8q8}, \href{https://go.wisc.edu/q827o9}{https://go.wisc.edu/q827o9}, \href{https://go.wisc.edu/83l84r}{https://go.wisc.edu/83l84r}.
The data were processed according to this script \href{https://go.wisc.edu/37x8hh}{https://go.wisc.edu/37x8hh}.
\end{itemize}

\subsection{Summary of MDSINE2 and FIDO}
\label{subsec:reference_methods}

This subsection briefly describes the MDSINE2 \citep{Bucci2016MDSINEMD, Gibson2021IntrinsicIO} and fido \citep{silverman2022} methods that were used in the simulation study.

\subsubsection*{MDSINE2}

MDSINE2 is a recently proposed Bayesian model of microbiome dynamics. It adapts the generalized Lotka-Volterra dynamics in the following ways,

\begin{enumerate}
	\item Taxa are assigned to clusters. This effectively reduces the dimensionality of the gLV's autoregressive dynamics -- taxa influence one another's growth rates via their cluster membership. Moreover, perturbation effects are constrained to be identical across all taxa within the same cluster.
	\item The approach is probabilistic and models each sample's total count and relative abundance structure. This contrasts with alternative gLV estimators, which often proceed by initially transforming abundance and applying regularized least squares.
\end{enumerate}
	
We provide a high-level overview of the model's generative mechanism. If we assume uniform sampling over time, then the model generates latent taxonomic abundances according to
\begin{align*}
	\log x_{j}^{(i)}\left(t + 1\right) \vert \mu_{j}^{(i)}\left(t\right) &\sim \mathcal{N}\left(\log\mu_{s,k}\left(t + 1\right), \sigma^2 \right)
\end{align*}
where $i, j$ and $t$ index subjects, taxa, and time. The mean vector $\mu_{j}^{(i)}$ is a deterministic, gLV-like function of random clustering and growth  parameters,
\begin{align*}
	\log \mu_{j}^{(i)}\left(t + 1\right) := &\log x_{j}^{(i)}\left(t\right) + a_{1,j}\left[1 + \sum_{p = 1}^{P} \gamma_{c_{j}}\mathbf{z}_{c_{j}, p}^{\left(\gamma\right)}\mathbf{1}\left\{(i, t) \in \texttt{Perturbation } p\right\}\right] - \\
	&a_{2,j}x_{j}^{(i)}\left(t\right) + \sum_{j': c_{j'} \neq c_{j}} b_{c_{j}c_{j'}}\mathbf{z}_{c_{j}c_{j'}}^{\left(b\right)}x_{j'}^{(i)}\left(t\right)
\end{align*}
The terms $a_{1,j}$ and $a_{2,j}$ are the growth and decay rates for taxon $j$, as in the standard gLV. 
$c_{j}$ is the cluster index of taxon $j$.
The summation of perturbations $p$ describes the influence of different perturbations on taxon $j$'s abundance. $\gamma_{c_{j}}$ and $\mathbf{z}_{c_{j}, p}^{\left(\gamma\right)}$ parameterized the strength and presence of a perturbation $p$ effect on cluster $c_{j}$ -- note that the perturbation influences are shared across all members of the same cluster. The final summation describes autoregressive dynamics between pairs of taxa $j, j'$. The autoregressive coefficients are shared between all pairs of taxa with the same cluster assignments $c_{j}, c_{j'}$. In this way, the autoregressive dynamics operate at the cluster level.

These latent taxonomic abundances $x_{j}^{(i)}\left(t\right)$ are transformed into the observed relative $y_{j}^{(i)}\left(t\right)$ and total $r_{i}\left(t\right)$ abundances for taxon $j$ in sample $t$ of subject $i$ using a negative binomial measurement model,
\begin{align*}
	y_{j}^{(i)}\left(t\right) \vert \left(x_{j}^{(i)}\left(t\right)\right)_{j = 1}^{J} \sim \text{NB}\left(\frac{r_{i}\left(t\right)x_{j}^{(i)}\left(t\right)}{\sum_{j'} x_{j'}^{(i)}\left(t\right)}, d_{1} + d_{0}\left(\frac{x_{j}^{(i)}\left(t\right)}{\sum_{j'} x_{j'}^{(i)}\left(t\right)}\right)^{-1} \right)
\end{align*}
where $d_{0}$ and $d_{1}$ are hyperparameters set in advance to account for dispersion in the observed samples.

Finally, priors are placed on the parameters $a_{1,j}, a_{2,j}, c_{j}, \gamma_{l}, \mathbf{z}^{(b)}_{l}, \mathbf{z}^{(\gamma)}_{l}$, and $b_{ll'}$. A stick-breaking process prior is placed on $c_{j}$, allowing for the number of clusters to adapt to the evidence in the data. Inference is performed through MCMC, cycling over these parameters and those used within hyperpriors.

\subsubsection*{fido}

fido combines a multinomial logistic-normal model, matrix-normal process, and Bayesian inference to model the effect of covariates on taxa abundance.

\begin{itemize}
    \item The multinomial proportions are transformed into real space. Then, the mean of the transformed data is assumed to be latent matrix-normal processes. This approach allows for the modeling of latent factors that capture the shared information across the taxa.

    \item The Bayesian framework starts with specifying priors for taxa covariance. Then, during each iteration of the collapse-uncollapse sampler, the \texttt{fido} updates the latent factors that capture the shared information across the taxa using latent - T process (LTP). In addition, the regression coefficients are updated that relate the latent factors and covariates, including perturbations.
\end{itemize}

Let's assume there are $J$ taxa, $Q$ covariates, and $N$ total samples. The model generative process for $k$-th taxa abundance at time $t$ in subject $i$, $y_{j}^{(i)}\left(t\right)$ is as follows. 
\begin{enumerate}
    \item The prior distribution of the covariance between additive log-ratio transformed (ALR) taxa is inverse Wishart distribution, $\Sigma_{J-1 \times J-1} \sim W^{-1}\left( \Xi, \upsilon\right)$ with a scale matrix $\Xi_{J-1 \times J-1}$ and degrees of freedom $\upsilon$.

    \item Then, the smooth mean function $\Lambda\left[X\right] \sim \text{GP}\left(\Theta\left[X\right],\Sigma, \Gamma\left[X\right] \right)$ relates the covariates $X_{Q \times N}$ to ALR-transformed $\eta$ with the mean function $\Theta\left[X\right]$, row (taxa) covariance $\Sigma$, and column (samples) covariance $\Gamma\left[X\right]$. 
        \begin{itemize}
        \item Covariates $X$ includes time $t$.
            \item $\Gamma\left[X\right]$ evaluates the kernel $K$ at time points $t$ and $t-1$ for a given subject $s$, $K\left(X_{s}\left(t\right), X_{s}\left(t-1\right) \right)$ otherwise, it is zero.
        \end{itemize}
    \item Next, $\eta \sim \text{N}\left(\Lambda\left[X\right], \Sigma, I_{N}\right)$ is a normal-matrix distribution, where  $ \pi = \phi^{-1}\left(\eta\right) $ and $\eta$ is a $(J-1) \times N$ real valued matrix.
    \item For the subject $i$ at time $t$, we compute the inverse of ALR, $\pi^{(i)}\left(t\right)$ and generate $\*y^{(i)}\left(t\right)$
    \begin{align*}
    \*y^{(i)}\left(t\right) \vert  \pi^{(i)}\left(t\right) \sim \text{Multinomial}\left(n^{(i)}\left(t\right),\pi^{(i)}\left(t\right)\right),
\end{align*} where $n^{(i)}\left(t\right)$ is the library size.
\end{enumerate}
	
\subsection{Supplementary Figures}

\begin{figure}
	\includegraphics[width=0.9\textwidth]{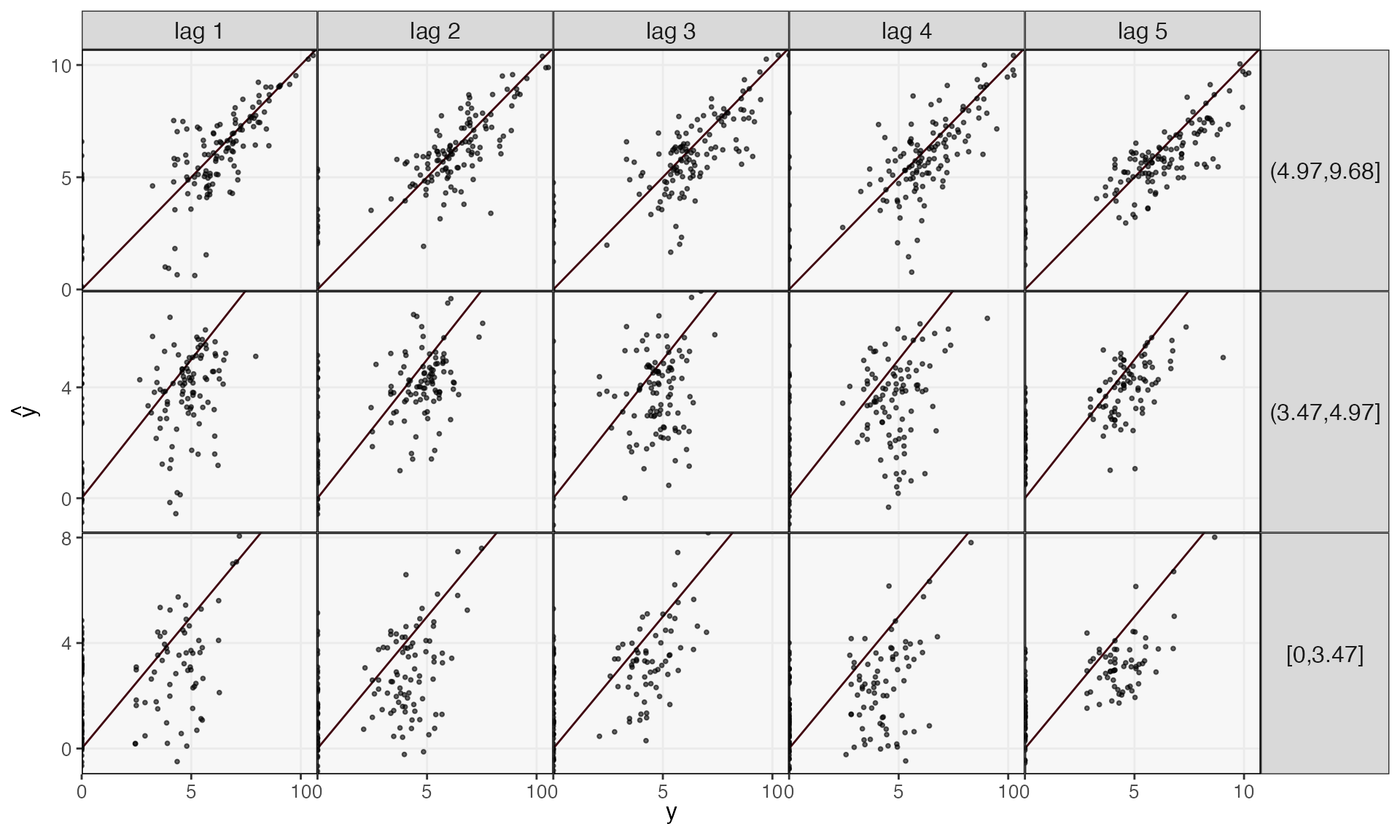}
	\label{fig:aqua_forecast_error}
	\caption{The analog of Figure 6 %\ref{fig:diet_forecast_error} 
 for the aquaculture case study in Section 4.3. %\ref{subsec:aquaculture}. 
 We only show in-sample errors, because our analysis only considers three unique tanks.}
\end{figure}

\begin{figure}
	\includegraphics[width=0.9\textwidth]{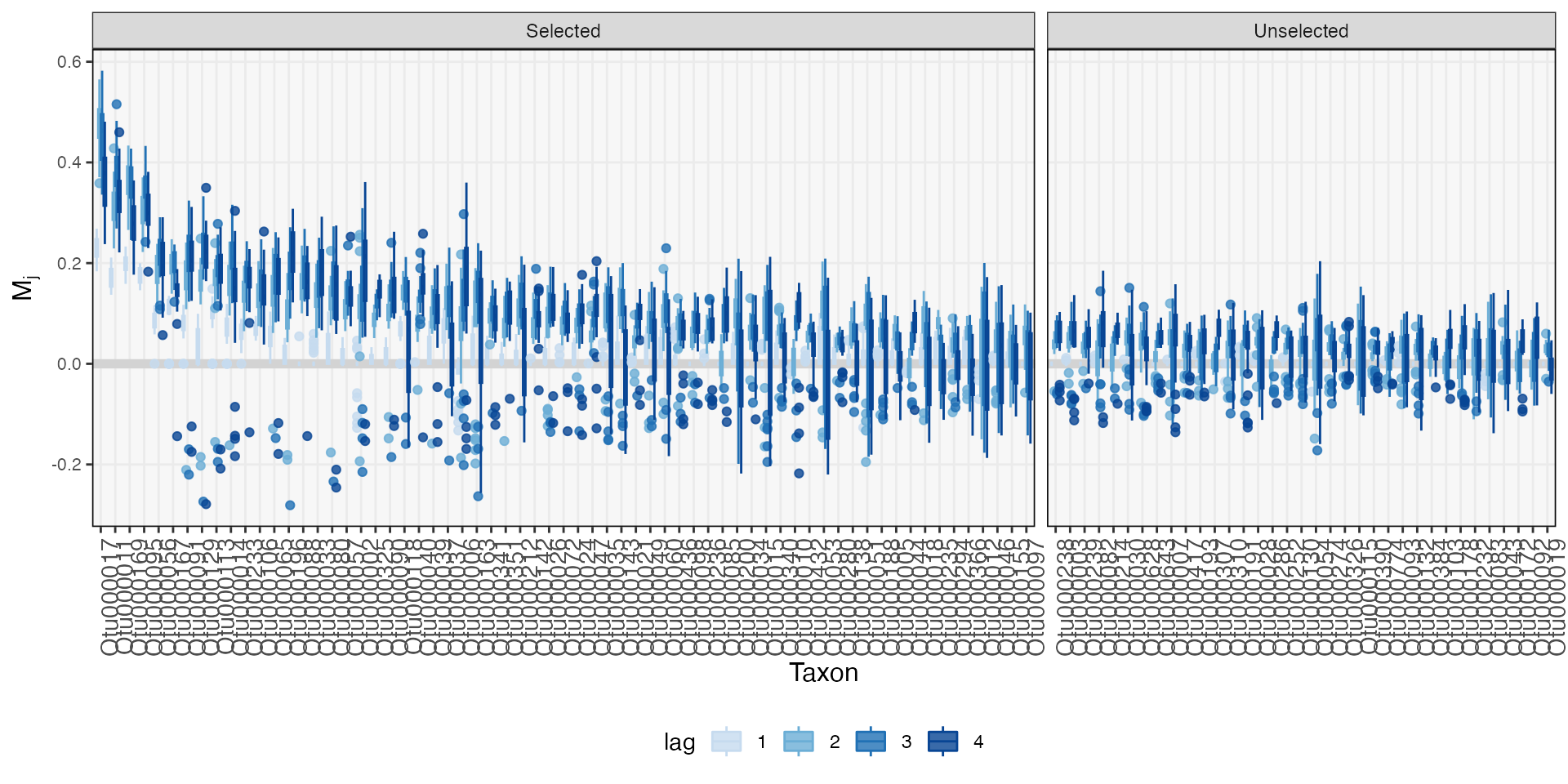}
	\caption{The distribution of mirror statistics $M_{j}$ for all selected and a subset of unselected taxa. Larger statistics indicate strong, consistent lag-0 effects (specifically, $PD_{j}\left(0\right)$ for taxon $j$) across data splits. The selection threshold is chosen adaptively according to a false discovery proportion estimate. The unselected taxa shown are those with the largest median $M_{j}$, and we have shown as many as possible while limiting the total number of boxplots to 100.}
\label{fig:diet_mirror_statistics}
\end{figure}

\begin{figure}
	\includegraphics[width=0.9\textwidth]{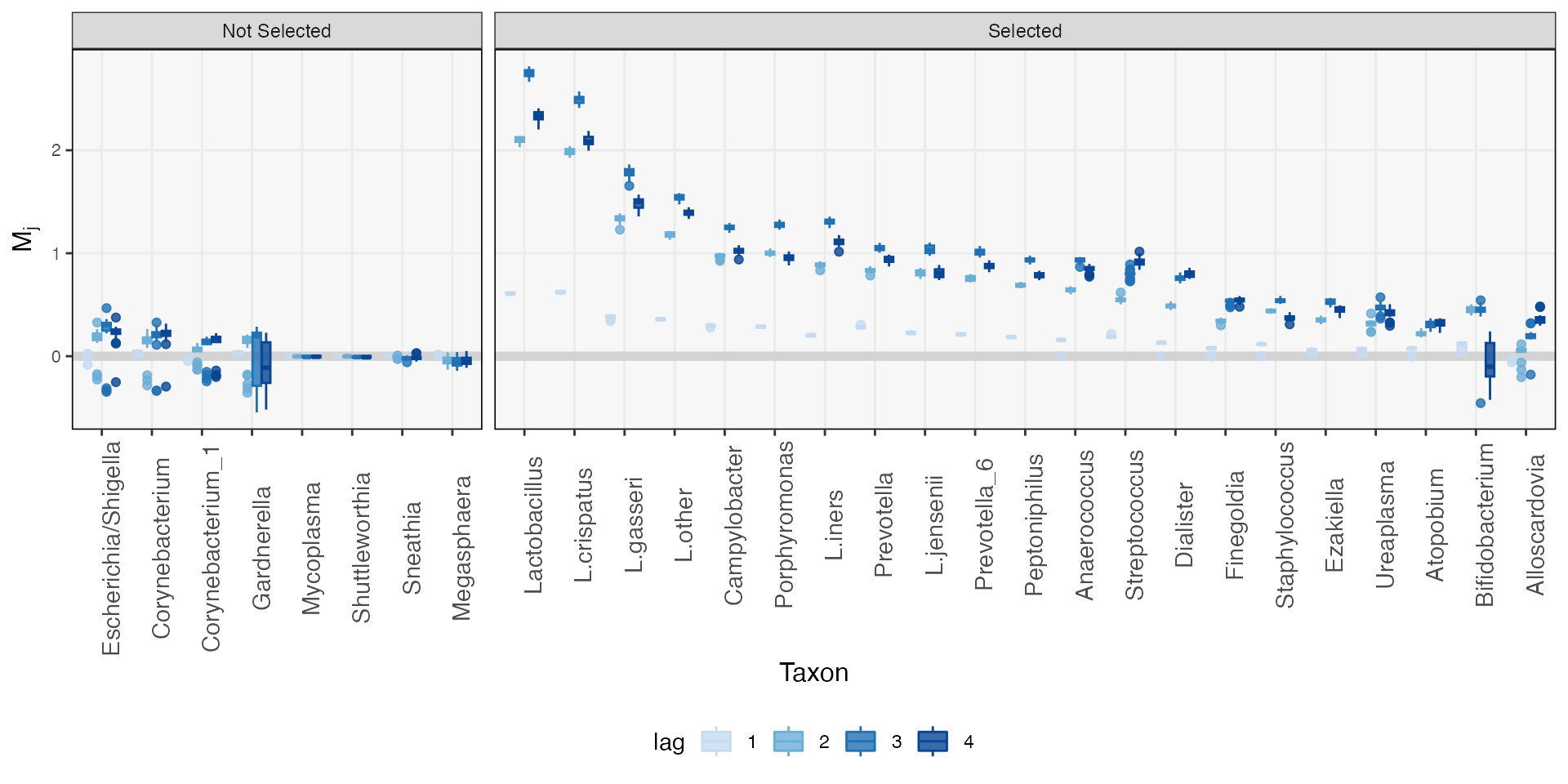}
	\caption{The analog of Supplementary Figure \ref{fig:diet_mirror_statistics}	 for the mirror statistics in the postpartum case study in Section 4.2. %\ref{subsec:postpartum}. 
 Mirror statistics appear more concentrated, likely a consequence of the larger sample size and stronger effects visible in this dataset.}
\label{fig:postpartum_mirror_statistics}
\end{figure}

\begin{figure}
	\includegraphics[width=0.9\textwidth]{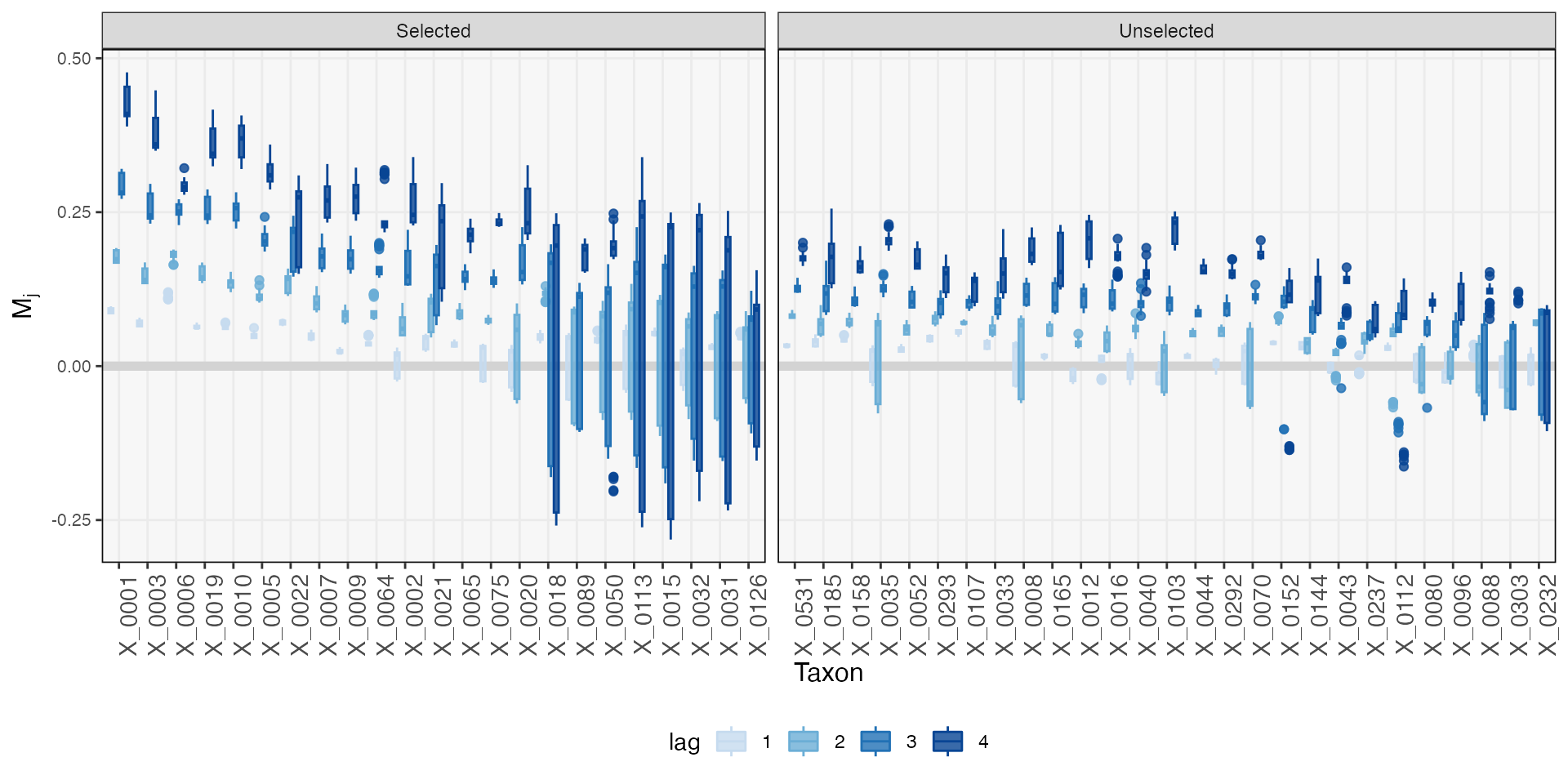}
	\caption{The analog of Supplementary Figure \ref{fig:diet_mirror_statistics} for the aquaculture case study in Section 4.3.%\ref{subsec:aquaculture}. 
 Note that we plot a taxon in the selected panel if it is significant for any lag. This explains why some taxa are selected despite having higher-lag mirror distributions that are symmetric around zero.}
\label{fig:aqua_mirror_statistics}
\end{figure}

%\bibliographystyle{biorefs}
%\bibliography{references.bib}
	
%\end{document}

\end{document}